\documentclass[fleqn,usenatbib]{mnras}

\usepackage{newtxtext,newtxmath}
\usepackage[T1]{fontenc}

\DeclareRobustCommand{\VAN}[3]{#2}
\let\VANthebibliography\thebibliography
\def\thebibliography{\DeclareRobustCommand{\VAN}[3]{##3}\VANthebibliography}

\usepackage{graphicx}	% Including figure files
\usepackage{amsmath}	% Advanced maths commands
\usepackage{pdflscape}	% Landscape pages

\newcommand{\eroextra}{eRO-ExTra}

\title[Nine supersoft TDEs in eROSITA-DE DR1]{Nine tidal disruption event candidates in eROSITA-DE DR1 discovered through supersoft X-ray selection}

\author[R. A. J. Eyles-Ferris et al.]{
R. A. J. Eyles-Ferris,$^{1}$\thanks{E-mail: raje1@leicester.ac.uk} R. L. C. Starling,$^{1}$ P. T. O'Brien,$^{1}$, K. L. Page$^{1}$ \& P. A. Evans$^{1}$
\\
$^{1}$School of Physics and Astronomy, University of Leicester, University Road, Leicester, LE1 7RH, UK
}

\date{Accepted XXX. Received YYY; in original form ZZZ}

\pubyear{\the\year}

\begin{document}
\label{firstpage}
\pagerange{\pageref{firstpage}--\pageref{lastpage}}
\maketitle

\begin{abstract}
Tidal disruption events are rare and diverse transients that occur when a star is torn apart by a supermassive black hole and accreted, which can result in a supersoft X-ray thermal transient. Here, we present nine tidal disruption event (TDE) candidates identified in eROSITA-DE Data Release 1 through a novel search for such supersoft sources. We select candidates by comparing the catalogued count rates in several combinations of bands and evaluate the nature of selected sources to produce our sample, among which five are entirely new X-ray TDE candidates. All our candidates' X-ray spectra are consistent with soft thermal emission and we show them to have faded through additional \textit{Swift} observations and catalogued data. We investigate publicly available data from ground- and space-based telescopes and find two of our sources have optical counterparts and four sources show flaring in their \textit{NEOWISE} IR light curves. The high proportion of our sources with IR flares compared to optically selected TDE samples could suggest a link between supersoft X-ray spectra and IR counterparts. We fit the IR light curves with a model of a spherical dust shell heated by the TDE and find these results to be broadly consistent with those of other TDEs with IR counterparts. Finally, we examine the host galaxies and show them to be similar to the general TDE host population.
\end{abstract}

\begin{keywords}
tidal disruption events -- catalogues -- surveys
\end{keywords}

%%%%%%%%%%%%%%%%%%%%%%%%%%%%%%%%%%%%%%%%%%%%%%%%%%

%%%%%%%%%%%%%%%%% BODY OF PAPER %%%%%%%%%%%%%%%%%%

\section{Introduction}

Tidal disruption events (TDEs) are astrophysical transients resulting from a star's self gravity being overcome by the tidal force of a supermassive black hole, torn apart and accreted \citep[e.g.][]{Rees88}. While the population of optically selected TDEs has grown steadily \citep[e.g.][]{Hammerstein23} and missions like \textit{Einstein Probe} \citep{Yuan22} and \textit{SVOM} promise to do the same for X-ray selected events, they remain rare. Despite the small  sample of $\sim$100 sources, the observed properties of TDEs are diverse. For instance, a significant fraction of events are detected only in X-rays or only in the optical, while other events are detected across the spectrum \citep[e.g.][]{Gezari21}. A small subsample also display the presence of relativistic jets \citep{Burrows11,Bloom11,Cenko12,Brown15,Pasham15,Andreoni22,Pasham23}, likely due to highly super-Eddington accretion at early times, while recent work has revealed IR counterparts to some TDEs as they heat the dust around the black hole \citep[e.g.][]{Lu16,vanVelzen16,Jiang21,Masterson24,Pasham25}. A number of TDEs have also been linked to quasi-periodic eruptors \citep[QPES, e.g.][]{Arcodia21,Miniutti23,Quintin23,Arcodia24,Nicholl24,Pasham25}, variable nuclear X-ray sources with similar spectral properties to TDEs.

In this work, we focus on X-ray bright TDEs. At X-ray energies, these are generally extremely soft thermal sources with temperatures of $\sim 100$ eV that decay slowly, often as $t^{-5/3}$ \citep[see][for a comprehensive review]{Saxton20}. X-ray catalogues offer a chance to expand transient populations and have been extensively used to identify new transient candidates and in particular TDEs \citep[e.g.][]{Esquej07,Saxton12,Saxton17,Saxton17a,Li20,Li22,EylesFerris22}. With the relatively small size of the TDE sample, expanding it is crucial to fully understanding the population. Generally, such searches compare the flux of serendipitously detected sources to catalogued data or limits to identify interesting candidates. Recently, however, searches focussing on sources with extremely soft spectra have also shown to be particularly effective at identifying TDEs \citep[][Eyles-Ferris et al. in prep.]{Sacchi23}.

The extended ROentgen Survey with an Imaging Telescope Array (eROSITA) is the main instrument aboard the \textit{Spectrum-Roentgen-Gamma} (\textit{SRG}) satellite. It consists of seven mirror assemblies and has a sensitivity roughly comparable to that of \textit{XMM-Newton}'s combined EPIC instruments at energies $\sim0.5 - 3$ keV but with a significantly greater field of view \citep{Predehl21}. \textit{SRG} was designed to continuously rotate with eROSITA in scan mode to build the eROSITA All Sky Survey (eRASS). The resulting data were collated every six months to produce a data release, referred to as eRASS 1, 2, 3 and so on. The survey was split in two with one galactic hemisphere belonging to the Russian consortium and the other belonging to the German consortium, eROSITA-DE. In January 2024, eROSITA-DE released the first six months of data covering December 2019 to June 2020 \citep[eRASS1, referred to hereafter as eROSITA-DE DR1,][]{Merloni24} and in this work, we apply supersoft spectral selection methods to this catalogue to identify TDE candidates. eROSITA-DE DR1 comprises both source catalogues and additional products including spectral files. Crucially, the source tables include count rates in various energy bands. By examining the relations between these count rates, we can determine a set of selection criteria that select only the softest sources. 

A TDE sample from the Russian half of the sky was presented by \citet{Sazonov21}. Those TDEs were found to be optically faint and their X-ray properties consistent with roughly Eddington levels of accretion onto the host supermassive black hole. There has also been work on identifying nuclear transients within eROSITA-DE DR1, in particular \citet{Grotova25} and \citet{Grotova25b}. In the first of those works, source variability was used to identify possible transients and various criteria were used to minimise AGN contamination, resulting in the \eroextra~catalogue. A sample of the 31 best TDE candidates from \eroextra~is presented in \citet{Grotova25b}, which also explores their multiwavelength and host properties. There is some crossover between the work presented here and this sample, however, our work includes additional data and analysis.

In Section \ref{sec:selection}, we discuss our count rate selection criteria and produce a sample of TDE candidates. We present our data analysis methodology and introduce additional data in Section \ref{sec:analysis} and apply them to constrain the properties of our candidates in Section \ref{sec:candidate_properties}. We briefly examine the properties of the host galaxies of our candidates in Section \ref{sec:host_properties} and summarise our conclusions in Section \ref{sec:conclusions}. Throughout this work, we assume a Planck cosmology \citep{Planck18} and give errors to 1-$\sigma$ confidence unless otherwise noted.

\section{Sample selection}
\label{sec:selection}

\subsection{Count rate selection criteria}

As mentioned above, eROSITA-DE DR1 provides count rates in different energy bands allowing us to use these to define our selection criteria. We compare the count rates of sources in the four lowest energy bands in the eROSITA-DE DR1 Main catalogue summarised in Table \ref{tab:bands}. The soft nature of thermal TDEs means the majority of the emission will be in this region of the spectrum with sufficient spectral variation to be differentiated from the general population. 

\begin{table}
    \caption{The eROSITA-DE DR1 energy bands used to define our selection criteria taken from Table D.2 of \citet{Merloni24}.}
    \centering
    \begin{tabular}{cc}
    \hline
     Band & Energy range (keV)\\
    \hline
    P1 & 0.2 -- 0.5\\
    P2 & 0.5 -- 1.0\\
    P3 & 1.0 -- 2.0\\
    P4 & 2.0 -- 5.0\\
    \hline
    \end{tabular}
    \label{tab:bands}
\end{table}

To ensure we select well-detected sources with high quality spectra, we first perform a signal to noise cut, requiring $SNR_{\rm P1} > 5$. Because the P1 band is particularly affected by absorption and even Galactic contributions could render it negligible, we also accept sources where $SNR_{\rm P1} > 3$ and $SNR_{\rm P2} > 5$. This yielded 13,853 sources ($\sim1.5$\% of the original catalogue).

\begin{figure*}
\centering
\includegraphics[width=0.42\textwidth]{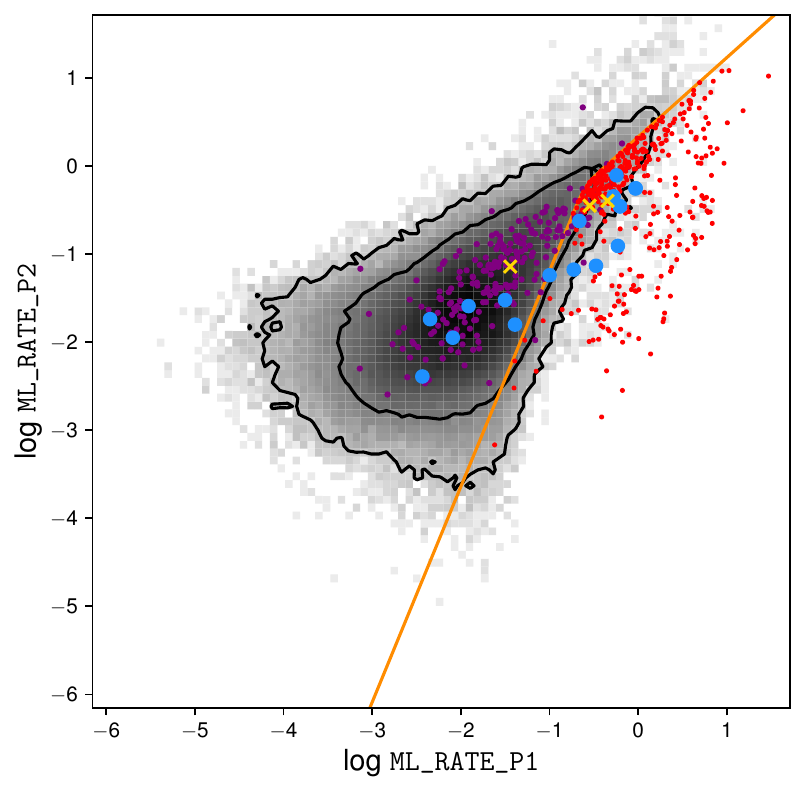}
\includegraphics[width=0.42\textwidth]{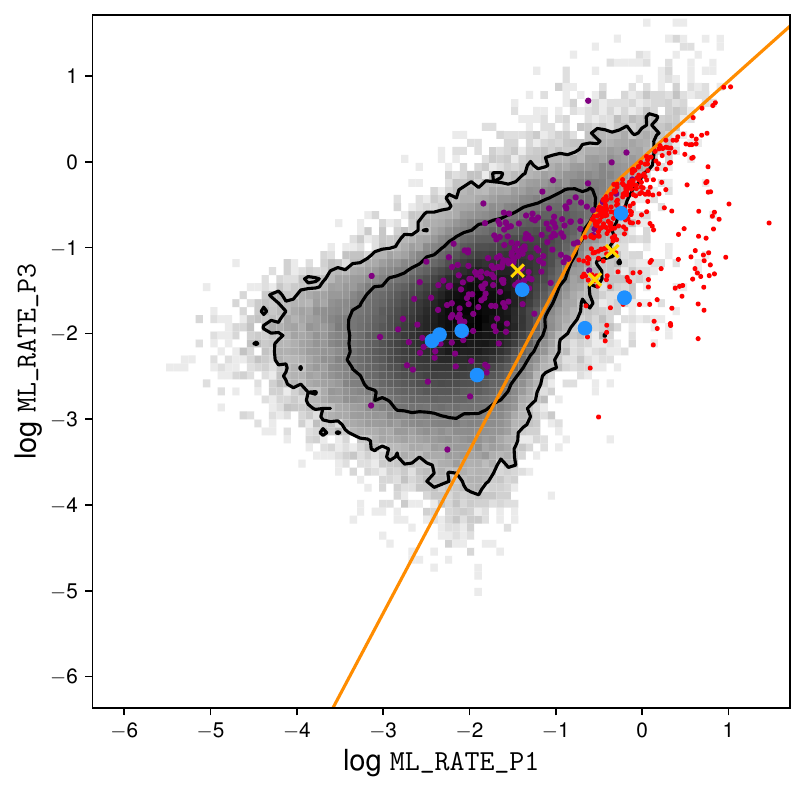}
\includegraphics[width=0.42\textwidth]{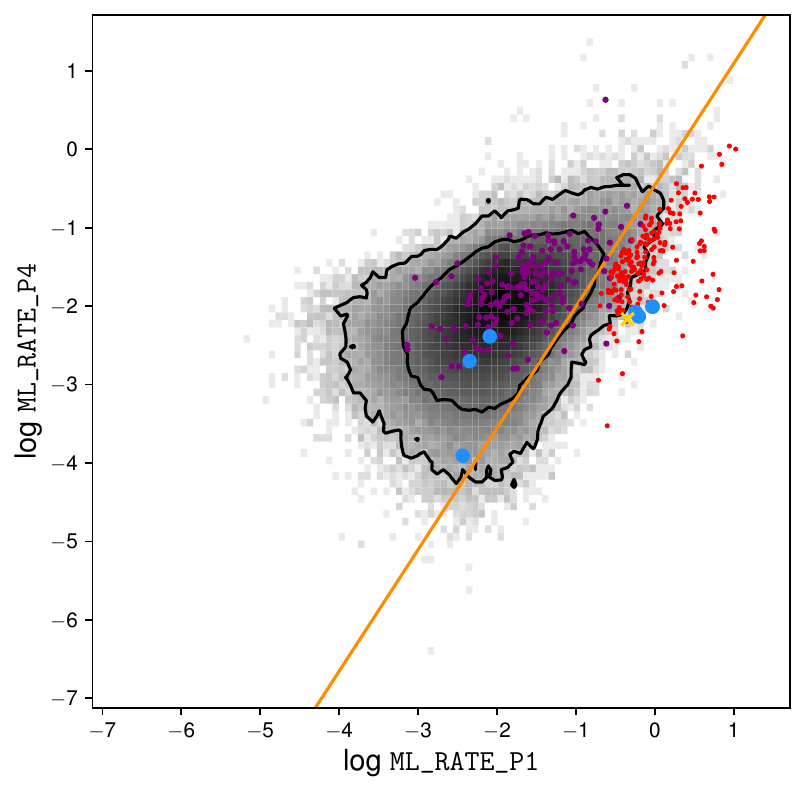}
\includegraphics[width=0.42\textwidth]{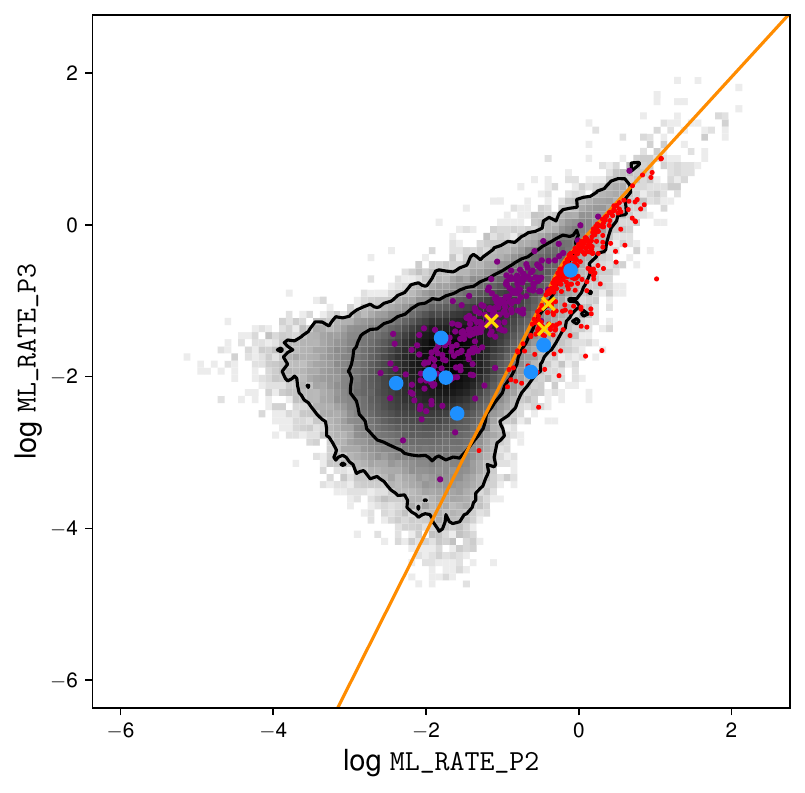}
\includegraphics[width=0.42\textwidth]{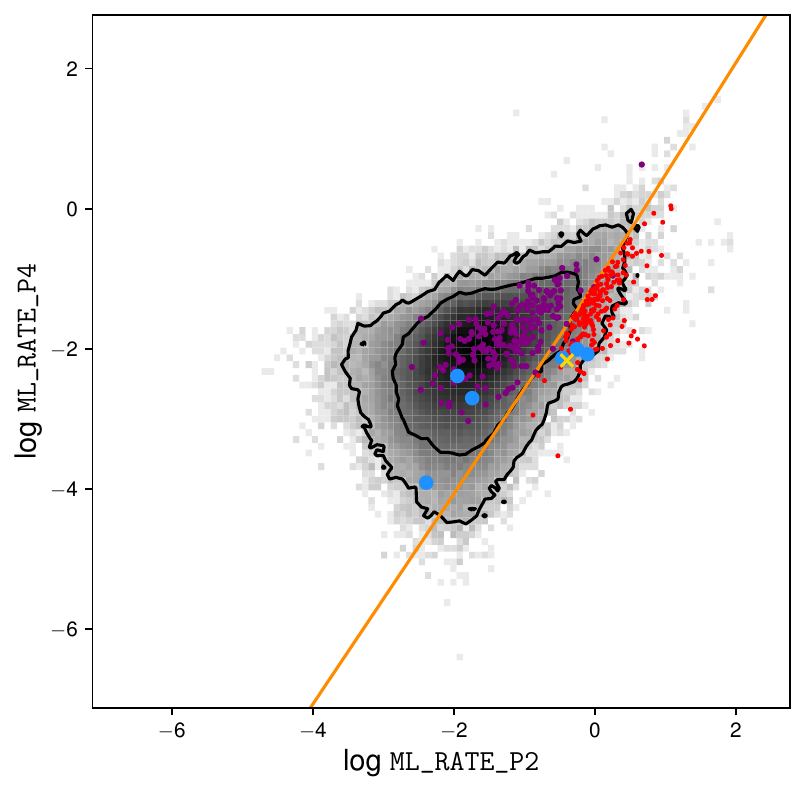}
\caption{The count rate cuts used to select sources for each combination of bands. The greyscale indicates the distribution of sources in log scale in the full eROSITA-DE DR1 Main catalogue with catalogued count rates in both plotted bands regardless of their SNR, the black lines are representative contours, the orange line indicates the cut function (see Equations \ref{eq:selection_p1p2} to \ref{eq:selection_p2p4}), blue points and gold crosses indicate known TDEs and QPEs respectively, the purple points are sources in the eRO-ExTra catalogue and the red points are sources selected by our cuts.}
\label{fig:cr_cuts}
\end{figure*}

Next, we define five further cuts based on the X-ray colours i.e. the ratios of the count rates in different bands. These aim to select the softest $\sim1$\% of the sources in the catalogue and we therefore inspected the distribution of X-ray colours of the full sample (Figure \ref{fig:cr_cuts}) We defined power laws, either singular or smoothly broken, which followed the contours of the distribution and are normalised to select approximately 1\% of the sources in the Main catalogue. The final criteria are plotted in Figure \ref{fig:cr_cuts} compared to the eROSITA-DE DR1 Main catalogue sources and are given by
\begin{equation}
    \scalebox{0.9}{$\texttt{ML\_RATE\_P2} < 0.4 \left(\frac{\texttt{ML\_RATE\_P1}}{0.25}\right)^{2.45} \left[\frac{1}{2}\left(1 + \left(\frac{\texttt{ML\_RATE\_P1}}{0.25}\right)^{10}\right)\right]^{-0.155}$}\,,
    \label{eq:selection_p1p2}
\end{equation}
\begin{equation}
    \scalebox{0.9}{$\texttt{ML\_RATE\_P3} < 0.9 \left(\frac{\texttt{ML\_RATE\_P1}}{0.55}\right)^{2.2} \left[\frac{1}{2}\left(1 + \left(\frac{\texttt{ML\_RATE\_P1}}{0.55}\right)^{10}\right)\right]^{-0.13}$}\,,
    \label{eq:selection_p1p3}
\end{equation}
\begin{equation}
    \scalebox{0.9}{$\texttt{ML\_RATE\_P4} < 0.1\,(\texttt{ML\_RATE\_P1})^{1.55}$}\,,
    \label{eq:selection_p1p4}
\end{equation}
\begin{equation}
    \scalebox{0.9}{$\texttt{ML\_RATE\_P3} < 0.3 \left(\frac{\texttt{ML\_RATE\_P2}}{0.6}\right)^{2} \left[\frac{1}{2}\left(1 + \left(\frac{\texttt{ML\_RATE\_P2}}{0.6}\right)^{10}\right)\right]^{-0.09}$}\,,
    \label{eq:selection_p2p3}
\end{equation}
and
\begin{equation}
    \scalebox{0.9}{$\texttt{ML\_RATE\_P4} < 0.2 \left(\frac{\texttt{ML\_RATE\_P2}}{2.7}\right)^{1.5} \left[\frac{1}{2}\left(1 + \left(\frac{\texttt{ML\_RATE\_P2}}{2.7}\right)^{10}\right)\right]^{0.01}$}\,.
    \label{eq:selection_p2p4}
\end{equation}
In Figure \ref{fig:cr_cuts}, we also highlight sources corresponding to the eRO-ExTra catalogue, the sample of \citet{Grotova25b}, QPE candidates eRO-QPE1, eRO-QPE2 and eRO-QPE3 \citep{Arcodia21,Arcodia24} and TDE candidates eRASSt J082337+042303 and eRASSt J043959-651405 identified via the \textit{Neil Gehrels Swift Observatory} (hereafter \textit{Swift}) Target of Opportunity (ToO) requests\footnote{\url{https://www.swift.psu.edu/toop/summary.php}} (if the source is detected in the appropriate bands). Our cuts would select the majority of the known TDEs and QPEs, although a significant fraction of \citet{Grotova25b}'s TDE sample are somewhat harder. The cuts result in 393 ($\sim 0.04$\%) of the Main catalogue sources being selected, of which 361 had $SNR_{\rm P1} > 5$. We highlight these sources in red in Figure \ref{fig:cr_cuts}.

We also apply our selection criteria to the eROSITA-DE Supplementary catalogue which has a lower \texttt{DET\_LIKELIHOOD\_0} threshold hence is more likely to include spurious sources. This check identified three additional sources, all with $SNR_{\rm P1} > 5$. Manual examination of these sources, however, indicate they are likely spurious detections (for instance, one is a detection towards the edge of another source's PSF) and we eliminate them at this stage.

Interestingly, Figure \ref{fig:cr_cuts} suggests that the sources in \eroextra~are broadly consistent with the general population of eROSITA-DE DR1 sources with no significant apparent bias towards supersoft spectra. This indicates that \eroextra~probes a different population to our work that may not be significantly dominated by typical soft thermal X-ray TDEs.

\subsection{Classification and external catalogue matches}

Because this sample is small, it is reasonable to perform a manual cross-check against other catalogues to broadly classify our sources. We use the Set of Identifications, Measurements and Bibliography for Astronomical Data \citep[SIMBAD,][]{Wenger00} database\footnote{\url{https://simbad.u-strasbg.fr/simbad/}} and the NASA/IPAC Extragalactic Database\footnote{\url{https://ned.ipac.caltech.edu/}} (NED), classifying the sources as previously known TDE or QPE candidates, active galactic nuclei (AGN), stellar sources, other sources not of interest in this search\footnote{This category included, for instance, X-ray binaries, cataclysmic binaries, novae, pulsars and a pair of interacting galaxies.} and finally, possible new TDE candidates. For this last classification, we also require the source to be spatially consistent within 3-$\sigma$ with the central position of a  possible host galaxy. The host in turn is required to display no previous evidence of AGN activity, for instance, using their \textit{Wide-field Infrared Survey Explorer} \citep[\textit{WISE,}][]{Wright10} colour using the AGN criteria of \citet{Stern12}, \citet{Mateos12}, \citet{Mingo16} and \citet{Mingo19} (see Section \ref{sec:host_properties} for further details). This also allows us to assign putative redshifts to three TDE candidates. We investigate the properties of the assumed host galaxies in Section \ref{sec:host_properties}.

\begin{table}
    \caption{Our classification of the source sample derived via our catalogue cuts.}
    \centering
    \begin{tabular}{cc}
    \hline
     Class & Number (percentage of sample)\\
    \hline
    Previously known TDES/QPEs & 5 (1.3\%) \\
    AGN & 133 (33.8\%) \\
    Stellar & 216 (55.0\%) \\
    Other & 30 (7.6\%) \\
    \hline
    TDE candidate & 9 (2.3\%) \\
    \hline
    \end{tabular}
    \label{tab:classification}
\end{table}

We summarise this classification in Table \ref{tab:classification}. The sample is dominated by stellar sources - this is expected as we had not eliminated Galactic sources prior to this stage. There are also a significant number of AGN, primarily Seyferts, which often have the soft thermal X-ray spectra we are targeting. As expected, we also identify the known TDEs and QPEs described above with the exception of eRO-QPE2. Finally, our criteria and this catalogue check did identify nine sources that appear to have both a soft X-ray spectrum and be spatially consistent with apparently inactive galaxies - properties that mark them as new candidate TDEs. Four of these sources are present in \eroextra~and the sample presented in \citet{Grotova25b}, however, the remaining five are new X-ray TDE candidates. Two additional sources from \citet{Grotova25b} were selected by our cuts, both of which were previously known optical transients. The crossover with \eroextra~also allows us to assign redshifts to an additional three candidates. A final search of the Transient Name Server\footnote{\url{https://www.wis-tns.org/}} finds 1eRASS J143819.6-261731 to be synonymous with AT2019tth, an unclassified transient identified by ZTF \citep{De19}. We present our sample in Table \ref{tab:sample}, including abbreviated names used throughout the rest of this paper. As mentioned above, while four sources are described in \citet{Grotova25b} (published after the submission of this work), we present additional data and analysis here and therefore include them in our sample.

\begin{table*}
    \caption{Our sample of supersoft TDE candidates, including redshifts, positions and whether they are also present in the \eroextra~catalogue.}
    \centering
    \begin{tabular}{ccccccc}
    \hline
     Source & Abbreviation & Date of eROSITA-DE DR1 detection & $z$ & RA & Dec & In \eroextra?\\
    \hline
1eRASS J030151.9-220017 & 1eRASS J0301 & 22 January 2020 & 0.0792$^1$ & 45.4665\degr & -22.0048\degr & N \\
1eRASS J075803.3+075526 & 1eRASS J0758 & 23 April 2020 & 0.0955$^2$ & 119.5138\degr & +7.9239\degr & N \\
1eRASS J091657.8+060955 & 1eRASS J0916 & 8 May 2020 & 0.091$^3$ & 139.2412\degr & +6.1655\degr & Y \\
1eRASS J103451.7-375057 & 1eRASS J1034 & 7 June 2020 & 0.06291$^4$ & 158.7155\degr & -37.8493\degr & N \\
1eRASS J143623.9-174103 & 1eRASS J1436 & 2 February 2020 & 0.1925$^3$ & 219.0999\degr & -17.6842\degr & Y \\
1eRASS J143819.6-261731 & 1eRASS J1438 & 6 February 2020 & 0.04732$^4$ & 219.5818\degr & -26.2921\degr & N \\
1eRASS J153406.2-090332 & 1eRASS J1534 & 21 February 2020 & 0.02404$^4$ & 233.5261\degr & -9.0590\degr & Y \\
1eRASS J190146.6-552200 & 1eRASS J1901 & 9 April 2020 & 0.059$^3$ & 285.4442\degr & -55.3669\degr & Y \\
1eRASS J222800.3-460514 & 1eRASS J2228 & 8 May 2020 & --- & 337.0015\degr & -46.0873\degr & N \\
    \hline
    \multicolumn{6}{l}{$^1$\citet{Gaia23}, $^2$\citet{Alam15}, $^3$\citet{Grotova25}, $^4$\citet{Jones09}.}\\
    \end{tabular}
    \label{tab:sample}
\end{table*}

\section{Data analysis}
\label{sec:analysis}

\subsection{X-ray analysis}

\subsubsection{Catalogued X-ray data}

We perform a wide X-ray catalogue search, finding no sources consistent with our candidates in the second \textit{ROSAT} all-sky survey catalogue \citep[2RXS,][]{Boller16}, the \textit{XMM-Newton} Serendipitous Source Catalogue \citep[4XMM-DR13 Version,][]{Webb20}, the \textit{Chandra} Source Catalog 2.0 \citep{Evans10} or the Living \textit{Swift} XRT Point Source Catalogue\footnote{Note that there is now a source corresponding to 1eRASS J0301 in LSXPS following our observations.} \citep[LSXPS,][]{Evans23}.

We also perform a search at the positions of each source using the HIgh-energy LIght-curve GeneraTor upper limit server \citep[HILIGT,][]{Saxton22,Konig22}, identifying \textit{ROSAT} upper limits for all the sources, albeit $\sim30$ years prior to the eROSITA detections. We also identify \textit{XMM-Newton} Slew Survey \citep{Saxton08} upper limits for four sources. However, one source, 1eRASS J0301, was detected in an \textit{XMM-Newton} slew in July 2020, approximately five months after its detection with eROSITA. While only consisting of a few ($\sim 9$) counts, this detection indicates the eROSITA observation could have occured while the source was still rising. We also searched the LSXPS upper limit server\footnote{\url{https://www.swift.ac.uk/LSXPS/ulserv.php}} but found that none of our candidates' positions had previously been covered by \textit{Swift}-XRT.

Finally, we examine the Monitor of All-sky X-ray Image \citep[MAXI,][]{Matsuoka09} observations of our source positions using the On-Demand service\footnote{\url{http://maxi.riken.jp/mxondem/}}. We cover the period of 58500 to 60000 MJD in 30 day bins using both the default Gas Slit Camera (GSC) bands and the Solid-state Slit Camera (SSC) at 0.5 to 2 keV. We find no significant evidence of any of our sources in the MAXI light curves, indicating them to be below the sensitivity limit.

\subsubsection{\textit{Swift} observations}

To examine the late time behaviour of our sources, around three to five years after their eROSITA-DE DR1 detection, we also obtained ToO observations with \textit{Swift}. For each source we requested an initial 2 ks observation. From the data obtained with \textit{Swift}'s X-ray Telescope (XRT), these observations resulted in non-detections to a 3-$\sigma$ significance for eight of our nine sources. We therefore use the LSXPS upper limit server \citep{Evans23} to calculate count rate upper limits at the source positions.

The remaining source, 1eRASS J0301, was detected in our initial \textit{Swift} observation in February 2024. We therefore obtained a further 5 ks observation in December 2024 as this allows us to constrain both the decay and the late time spectrum of the source. We acquired the processed data and spectral files for this source using the User Objects service available from the UK \textit{Swift} Science Data Centre\footnote{\url{https://www.swift.ac.uk/user_objects/}} \citep{Evans09}.

\subsubsection{Spectral fitting procedure}

To better characterise our sample, we use \textsc{Xspec v12.13.1} \citep{xspec} to fit their X-ray spectra. We obtained the eROSITA spectral files from the eROSITA-DE DR1 Data Archive (eRODat)\footnote{\url{https://erosita.mpe.mpg.de/dr1/erodat/}}. The spectra were grouped using the \textsc{ftgrouppha} utility from \textsc{FTOOLS} \citep{ftools} to ensure each bin had a minimum of one count allowing fitting with Cash-statistics and we ignore photons above 2.3 keV - in most cases, the count rate above this was insignificant\footnote{Note that this is unsurprising and a natural effect of our selection criteria.} and including these photons at higher energies made negligible difference to the results of our fits. For the \textit{Swift} spectrum of 1eRASS J0301 obtained in December 2024, we also ignore photons above 2.3 keV and fit the data using the same procedure.

We include Galactic absorption using \texttt{tbabs} with column densities taken from the maps of \citet{Willingale13} and a host contribution using \texttt{ztbabs} for sources where we had a host redshift. We initially examine two models that are redshifted if a host redshift is available, a blackbody and a power law. We then calculate the luminosity based on the blackbody model. For 1eRASS J2228 which has no known redshift, we assumed $z=0.1$, approximately commensurate with the peak of the TDE redshift distribution \citep[e.g.][]{Qin22} and compatible with the putative redshifts of the other sources in the sample. We also examine a combined blackbody + power law model but find that it doesn't significantly improve our fits and instead, many of the sources converge to either the blackbody or power law model. For the eRASS 2 - 5, i.e. the second through to fifth epochs of the eROSITA All Sky Survey, data available from \eroextra~and most of the upper limits obtained from the Upper Limit Server and \textit{Swift}, we assume the same model as the eROSITA-DE DR1 data. The exception is the first \textit{Swift} observation of 1eRASS J0301 where we assume the same model as the December 2024 \textit{Swift} observation. We then derive fluxes and luminosities from the limits using \textsc{PIMMS v4.12a}\footnote{Note that for the \eroextra~eRASS 2 - 5 data, we first convert the tabulated flux to a count rate based on \citet{Grotova25}'s fitted model before deriving the luminosity assuming a blackbody model.}. The results of our spectral fitting are given in Table \ref{tab:spectra} and Figure \ref{fig:spectra} and discussed in Section \ref{sec:candidate_spectra}.

\subsection{Multiwavelength data}

To further characterise our candidates, we also explore their multiwavelength properties. Some of these data come from our \textit{Swift} ToOs and we add to these by identifying counterparts to our sources in publicly available data from ground based observatories.

\subsubsection{\textit{Swift}/UVOT ultraviolet photometry}

\textit{Swift}, in addition to the XRT, observes simultaneously with its Ultraviolet/Optical Telescope (UVOT). We therefore also analyse these data to constrain the lower energy behaviour of our sources. All of our initial observations used the \textit{uvm2} filter (central wavelength 224.6 nm) while the 5 ks observation of 1eRASS J0301 used the \textit{uvw1} filter (central wavelength 260.0 nm).

We use the \textsc{uvotproduct v2.9} to measure the photometry of our sources using a 5 arcsecond radius aperture centred on the eROSITA position. We use a 5-$\sigma$ threshold unless otherwise noted and where a source is not detected, calculate the 3-$\sigma$ upper limit in the aperture. In the case of 1eRASS J1534, \textsc{uvotproduct} is unable to measure the flux due to proximity to a very bright source. We therefore use \textsc{uvotsource v4.5} and a manually defined background region to obtain upper limits for this source although we note there are still likely to be inaccuracies in the photometry. We summarise the results of the UVOT observations in Table \ref{tab:uvot_photometry}. Note that several of the \textit{Swift} observations were split across multiple windows due to interruptions and we therefore derive UVOT photometry for each window. We find we detect a UV counterpart for six of our sources. Due to the large separation in time since the presumed peak in our sources' light curves, these are likely to be dominated by host emission but may have a lingering contribution from the TDE \citep[e.g.][]{Mummery24a,Mummery24b}.

\subsubsection{Optical photometry}

\begin{figure*}
    \centering
    \includegraphics[width=\columnwidth]{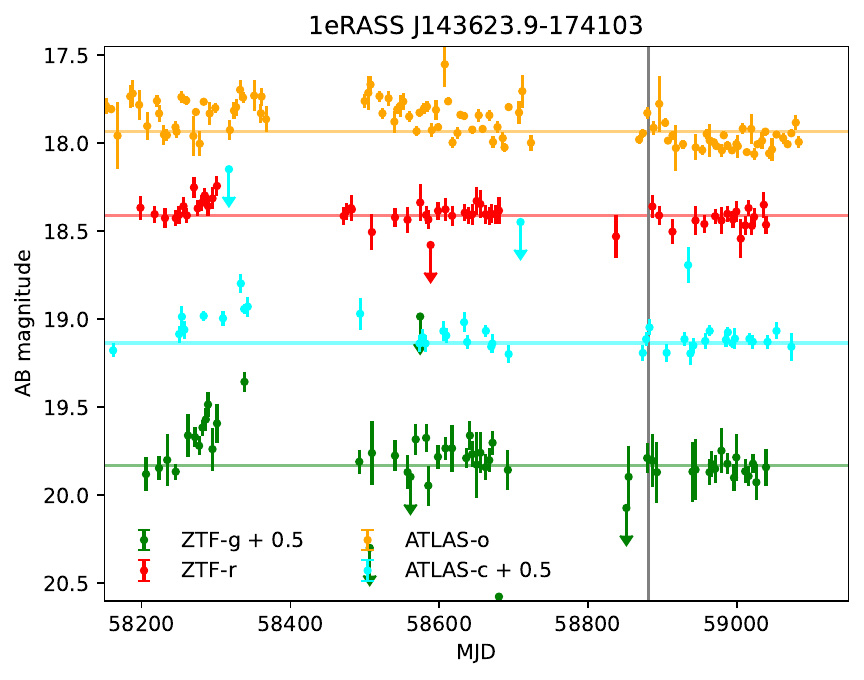}
    \includegraphics[width=\columnwidth]{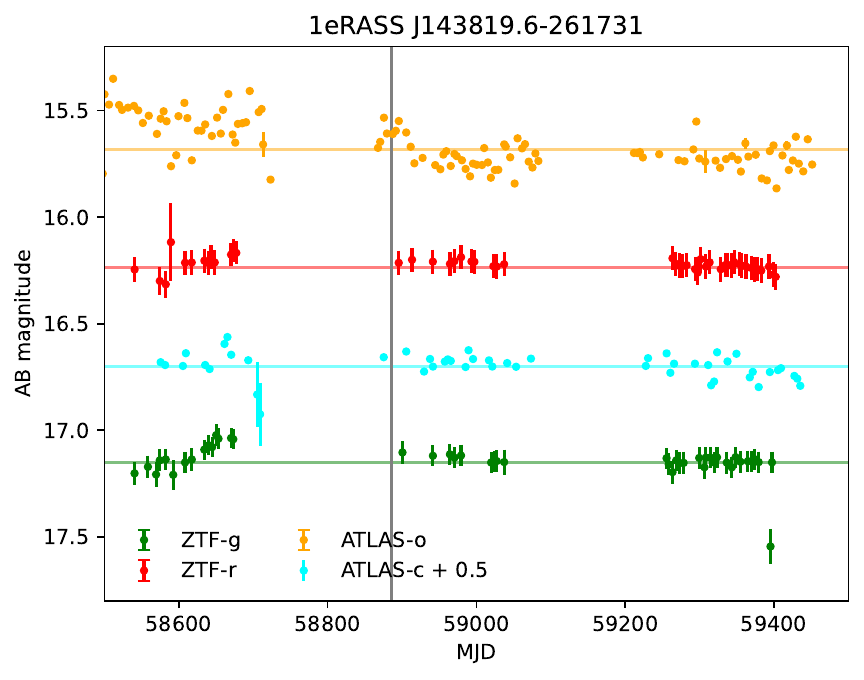}
    \caption{The optical light curves of 1eRASS J1436 and 1eRASS J1438 obtained from ZTF and ATLAS forced photometry. In both panels, errors are given to 1-$\sigma$, the grey vertical line indicates the time of the eROSITA-DE DR1 detections and the horizontal lines indicate the mean magnitude in each filter to aid identification of possible flares.}
    \label{fig:opt_lcs}
\end{figure*}

We examine the photometry available from the Zwicky Transient Facility \citep[ZTF,][]{Masci19}. ZTF observes the visible sky in several filters corresponding to the Sloan Digital Sky Survey (SDSS) or Pan-STARRs broadband filters. First we utilise the photometry from Data Release 22 (DR22) available via the NASA/IPAC Infrared Science Archive\footnote{\url{https://irsa.ipac.caltech.edu/Missions/ztf.html}}. 

We crossmatch to the DR22 source catalogue with a search radius of 5\arcsec. We find seven sources have counterparts with 1eRASS J1901 and 1eRASS J2228 being too far South and therefore not in the footprint of the survey. These counterparts will be dominated by host emission, therefore we looked for variability in the archival light curves as an indicator of TDE emission. We find that two of the seven sources with counterparts do show low significance signs of variability in both the \textit{g} and \textit{r} bands they were observed in. One of these is 1eRASS J1438 which is synonymous with the ZTF detected AT 2019tth and therefore an optical flare is to be expected. However, the counterpart to 1eRASS J1436 also shows evidence of variability in its light curve. Notably, 1eRASS J0301 does not show significant variation in its light curve despite continued bright X-ray emission, strengthening its position as an X-ray only TDE.

For the two sources with possible optical variability, we take the \textit{g} band positions from DR22\footnote{We did perform image subtraction to attempt to determine the position of these counterparts more accurately but were unable to obtain a subtraction with sufficient quality. In the case of 1eRASS J1438/AT 2019tth, the position we use is consistent to $<0.1$\arcsec~with the position reported by \citet{De19}.} and use the ZTF forced-photometry service \citep{Masci23} to measure the flux through PSF fitting centred on those positions. These fluxes are rebinned to approximately 5 day bins to increase signal to noise and converted to calibrated AB magnitudes using the procedure outlined in the forced-photometry documentation\footnote{\url{https://irsa.ipac.caltech.edu/data/ZTF/docs/ztf_forced_photometry.pdf}} with a 5-$\sigma$ detection threshold and 3-$\sigma$ upper limits.

We also used the Asteroid Terrestrial-impact Last Alert System forced-photometry service\footnote{\url{https://fallingstar-data.com/forcedphot/}} \citep[ATLAS,][]{Tonry18,Smith20,Shingles21} to further expand our optical photometry. ATLAS is also a wide-field imaging observatory with four nodes across Hawaii, Chile and South Africa and images the entire visible sky at a high cadence using broad band cyan (\textit{c}, 420 -- 650 nm) and orange (\textit{o}, 560 -- 820 nm) filters. The forced-photometry service derives a PSF model for a given image and fits it to a user provided fixed position - in this case we use the ZTF positions as above for consistency - to measure the flux at that position. We further process these data to rebin into 5 day bins and limit our light curves to 5-$\sigma$ detections or 3-$\sigma$ upper limits. We find no evidence of optical flaring in most of our sources for several hundred days before and after the eROSITA detection, however, similarly to the ZTF results, 1eRASS J1436 and 1eRASS J1438 display some low signficance variability. We do note that the eROSITA detection of 1eRASS J2228 occurred during a gap in the light curve, but again there was little evidence of flaring before or after this gap. 

We show the full optical light curves of 1eRASS J1436 and 1eRASS J1438 in Figure \ref{fig:opt_lcs}. Due to gaps in coverage, it is difficult to constrain the peak times of these events, however, by eye we estimate the peaks to be around MJDs $\sim58350$ and $\sim58700$ (with errors of order 50 -- 100 days) for 1eRASS J1436 and 1eRASS J1438 respectively. In both cases this is hundreds of days prior to the eROSITA detection. The gap in coverage also means the post-peak evolution of the sources is unobserved - by the time ZTF and ATLAS next observed they had essentially plateaued. Thus, while we conclude that optical flares likely did accompany the X-ray emission, and possibly preceded them, we cannot constrain their properties further.

\subsubsection{\textit{NEOWISE} IR photometry}

Finally, we examine the IR properties of our sources using the Final Data Release of the \textit{Near-Earth Object Wide-field Infrared Survey Explorer Reactivation Mission} \citep[\textit{NEOWISE},][]{Mainzer14}. \textit{NEOWISE} has performed all-sky infrared surveys at 3.4 $\mu$m (\textit{W1}) and 4.6 $\mu$m (\textit{W2}) from December 2013 to August 2024. We again use the NASA/IPAC Infrared Science Archive\footnote{\url{https://irsa.ipac.caltech.edu/Missions/wise.html}} to acquire the relevant data for our TDE candidates.

To identify IR counterparts, we use a custom pipeline. A full description of this pipeline will be given in a future work but to summarise, individual exposures are downloaded from the Simple Image Access v2\footnote{\url{https://irsa.ipac.caltech.edu/ibe/sia.html}} service, cosmic rays cleaned using the LACosmic Laplacian edge detection algorithm \citep{vanDokkum01} and the exposures aligned and stacked using the \texttt{astroalign v2.6.1} module. Image subtraction is performed using the ZOGY algorithm \citep{Zackay16} as implemented by \texttt{PyZOGY} \citep{Guevel21} and template images taken from the unWISE coadded AllWISE images \citep{Lang14,Meisner17a,Meisner17b}. Any residual flux is measured using PSF photometry calibrated to the AllWISE Source Catalogue \citep{Cutri13}. We set a detection threshold of 5-$\sigma$ and find four sources exhibit significant flaring, 1eRASS J0758, 1eRASS J1436, 1eRASS J1438 and 1eRASS J2228. We present and analyse the light curves of those sources in Section \ref{sec:ir_behaviour}.

\section{TDE candidate properties}
\label{sec:candidate_properties}

\subsection{X-ray properties}

\subsubsection{Spectral properties}
\label{sec:candidate_spectra}

We show the results of our X-ray spectral fitting in Table \ref{tab:spectra} and Figure \ref{fig:spectra}. We find that six of the sources are statistically best fit with absorbed blackbody spectra of temperatures $\sim 50$ to $\sim 130$ eV, consistent with expectations for TDEs. The remaining three sources are statistically best fit with absorbed power law models, however, with photon indices of $4 < \Gamma < 6$. Such steep spectra are atypical of AGN or other nuclear power law sources and instead it is likely that the blackbody model more accurately captures these objects behaviour assuming the absorption to be accurately fitted. The luminosities given in Table \ref{tab:spectra} are therefore derived using the blackbody fit. Five of our sources are also significantly brighter than both earlier limits, although these are all derived from \textit{ROSAT} survey or pointed observations and are therefore $\sim30$ years prior to the eROSITA detections, and the results of our \textit{Swift} observations. We show the resulting light curves in Figures \ref{fig:xray_lcs_fitted} and \ref{fig:xray_lcs}.

\begin{figure}
    \centering
    \includegraphics[width=\columnwidth]{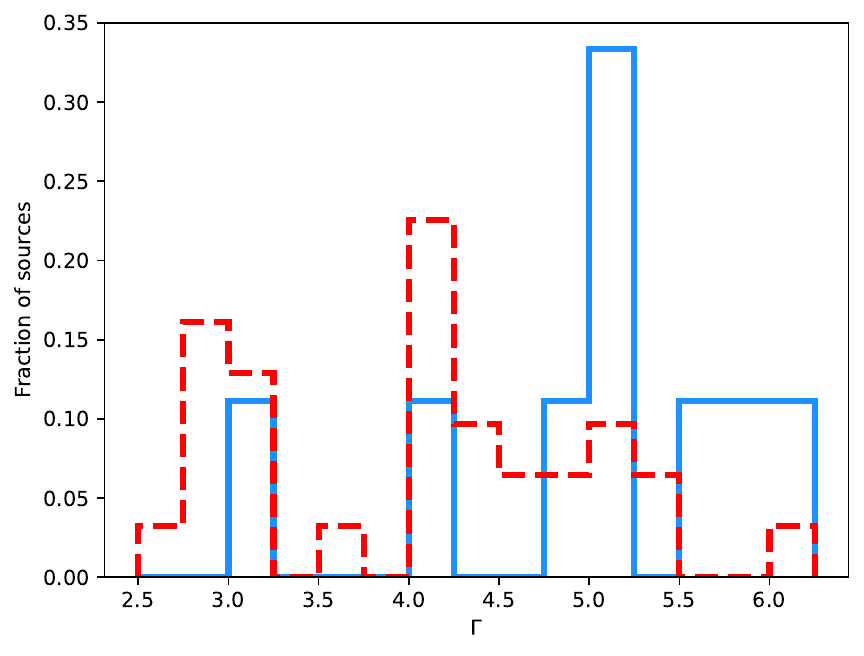}
    \caption{The photon index distribution of our TDE candidates (blue) compared to the sample of \citet{Grotova25b} (red dashed).}
    \label{fig:gamma_comp}
\end{figure}

We find our sample is generally spectrally softer than that examined in \citet{Grotova25b}, as shown in Figure \ref{fig:gamma_comp}, attributable to our selection criteria which select only the softest sources. For the four sources in common between our samples, our values of $\Gamma$ are all consistent with those found by \citet{Grotova25b} with minor discrepancies likely arising from us also incorporating host absorption.

We also examine the spectral variability of 1eRASS J0301 as this source remains X-ray bright at later times. Enough photons have therefore been accumulated ($>40$ counts) in our \textit{Swift} observations to allow us to fit the spectrum to a reasonable accuracy. Interestingly, this source appears to harden and the fit to the later XRT spectrum favours the absorbed power law model, as shown in Table \ref{tab:spectra}. The reason for this is unclear but could be due to transitioning to a harder state as has been observed in several other TDEs \citep[e.g.][]{Komossa04b,Wevers21,Lin22} and in a similar way to X-ray binaries \citep[e.g.][]{Mummery21,Mummery23}.

Overall, the spectral properties of these sources are entirely consistent with the wider population of soft X-ray TDEs which often display blackbody temperatures of order 50 -- 100 eV and luminosities of $10^{42}$ -- $10^{44}$ erg s$^{-1}$ as we find here \citep[e.g.][]{Saxton20}. The black hole masses we find below (see Section \ref{sec:host_properties}) imply the brightest measured X-ray luminosities of our sources are $\sim0.003$ to $0.6$ of the Eddington luminosity. The true peak luminosities are likely to have somewhat larger Eddington ratios but our results are consistent with the general result that accretion rates are a significant fraction of Eddington.

\subsubsection{Temporal properties}

\begin{figure*}
\centering
\includegraphics[width=0.45\textwidth]{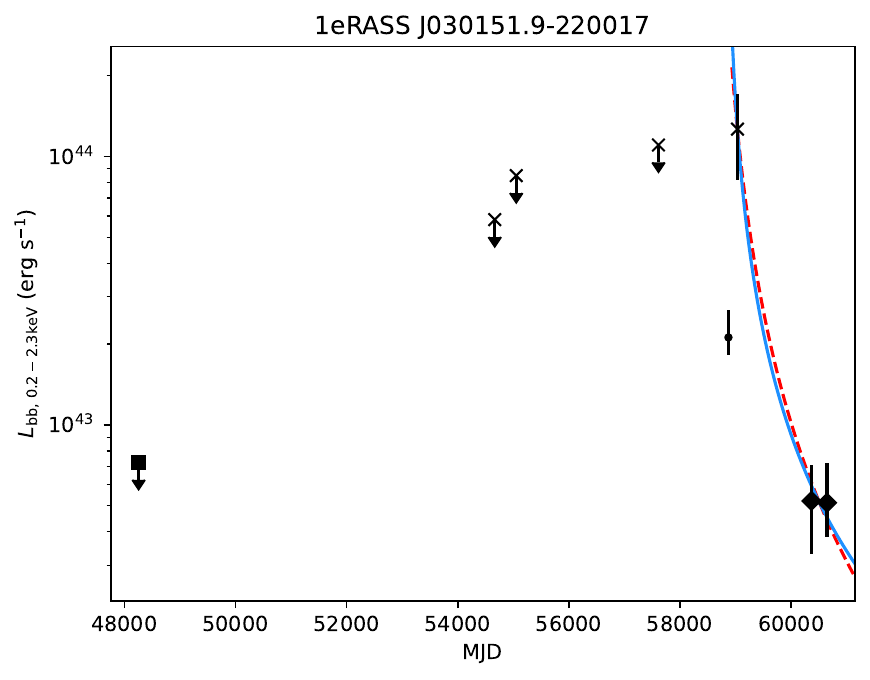}
\includegraphics[width=0.45\textwidth]{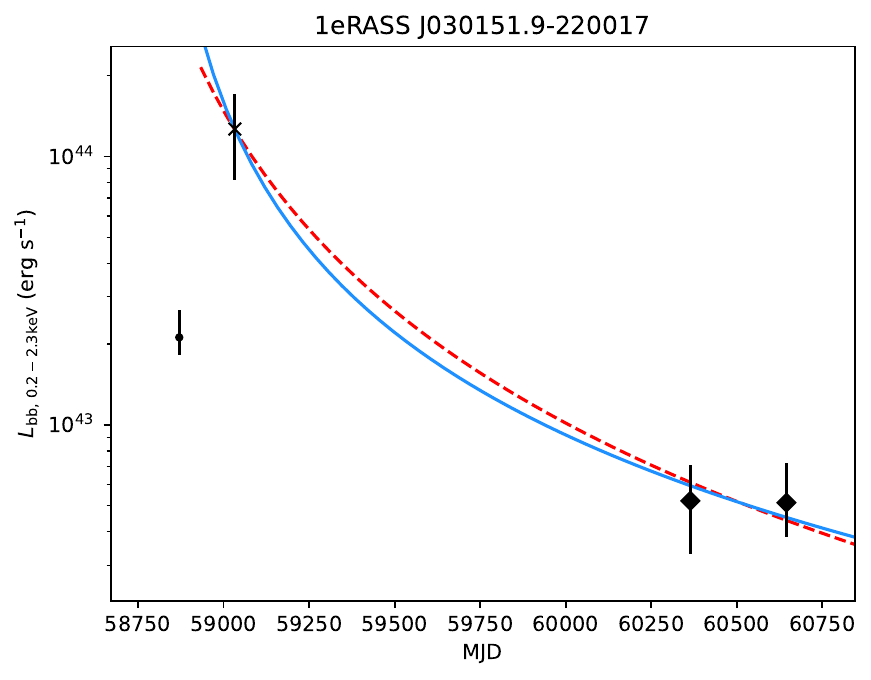}
\includegraphics[width=0.45\textwidth]{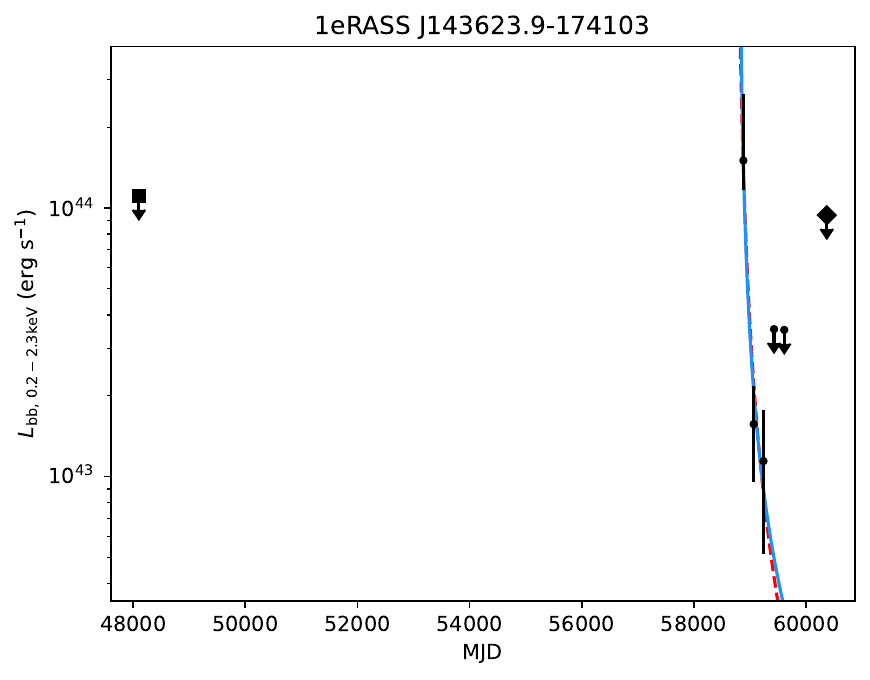}
\includegraphics[width=0.45\textwidth]{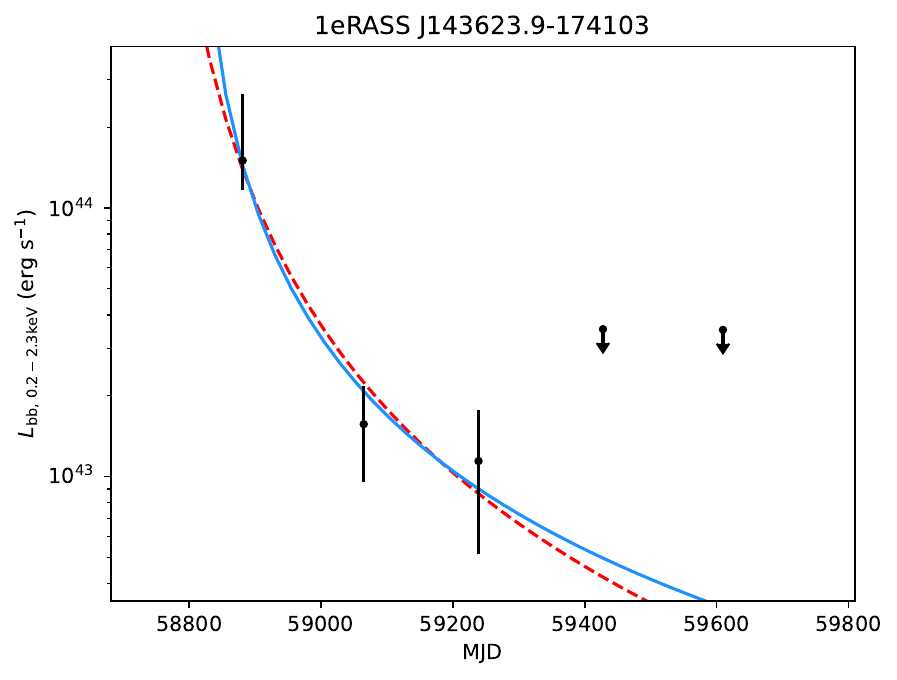}
\caption{The 0.2-2.3 keV X-ray luminosity light curves for 1eRASS J0301 (top row) and 1eRASS J1436 (bottom row). The left column shows the complete light curve and the right column shows a zoom-in of the period dominated by the TDE. Data from eROSITA, \textit{ROSAT}, \textit{XMM} and \textit{Swift} are marked with circles, squares, crosses and diamonds respectively. We plot fits to the light curves assuming full ($t^{-5/3}$) and partial ($t^{-9/4}$)  disruptions are in solid blue and dashed red lines respectively.}
\label{fig:xray_lcs_fitted}
\end{figure*}

 \begin{figure*}
\centering
\includegraphics[width=0.4\textwidth]{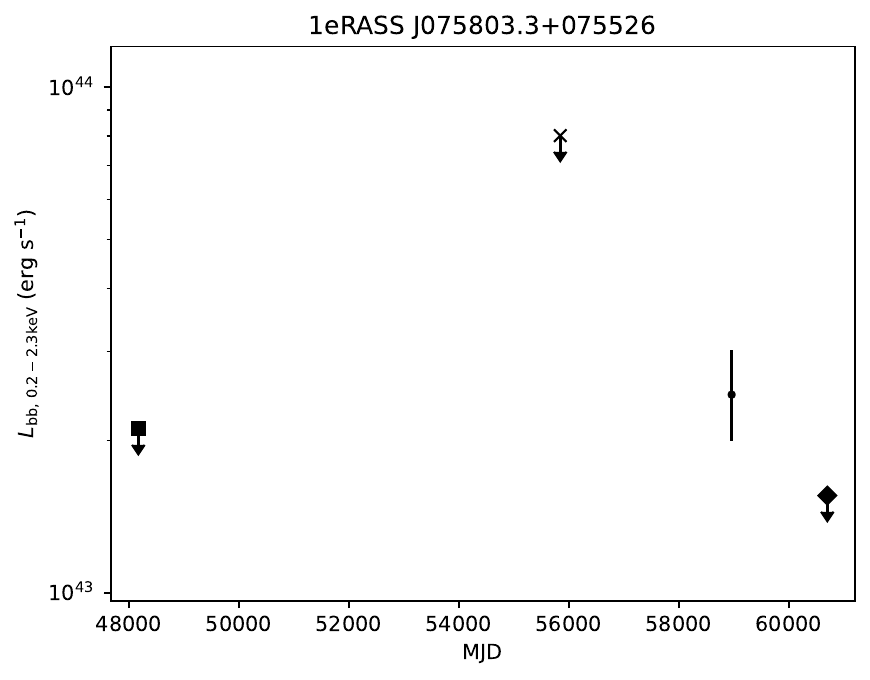}
\includegraphics[width=0.4\textwidth]{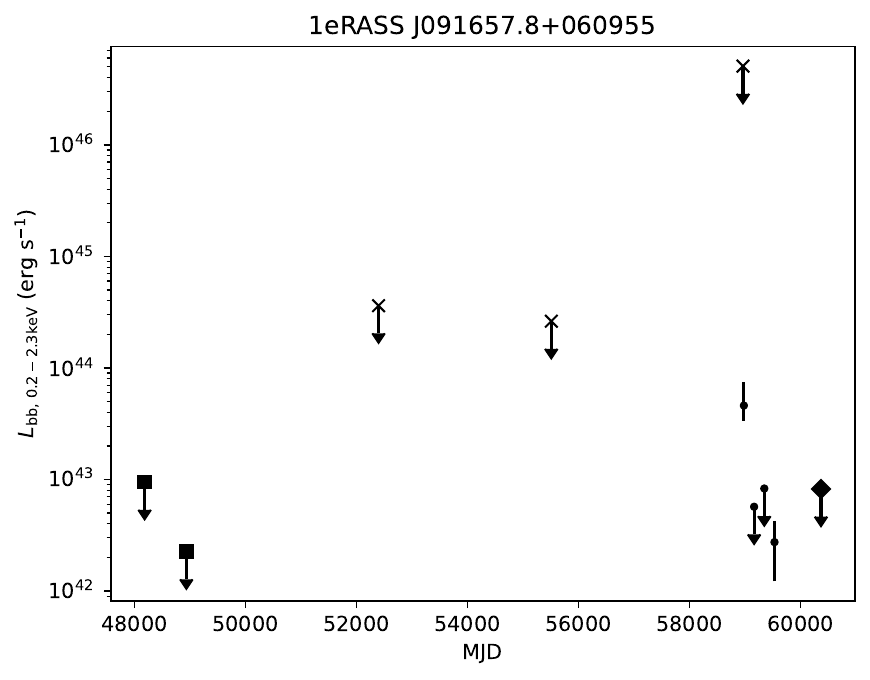}
\includegraphics[width=0.4\textwidth]{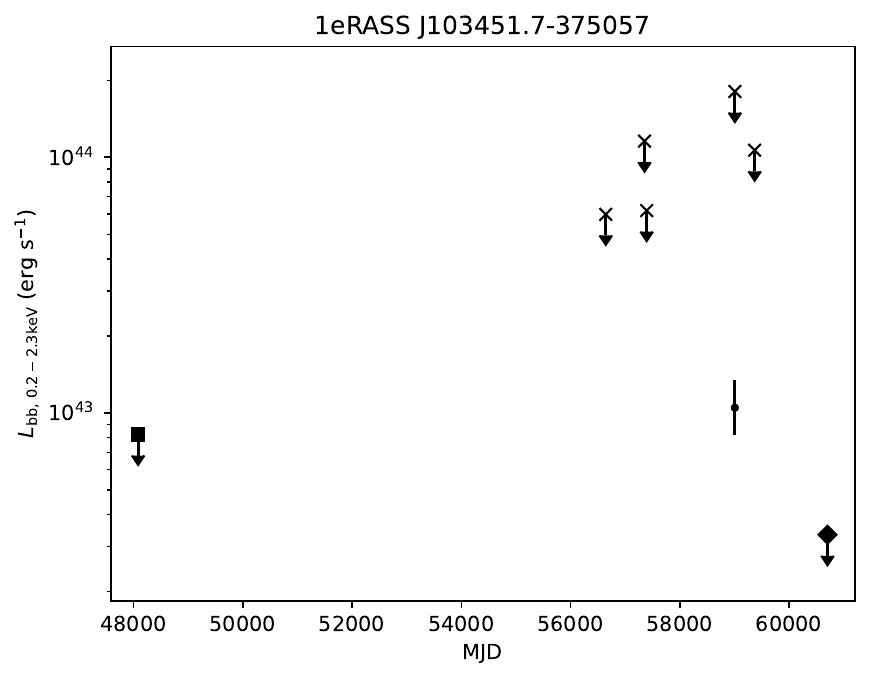}
\includegraphics[width=0.4\textwidth]{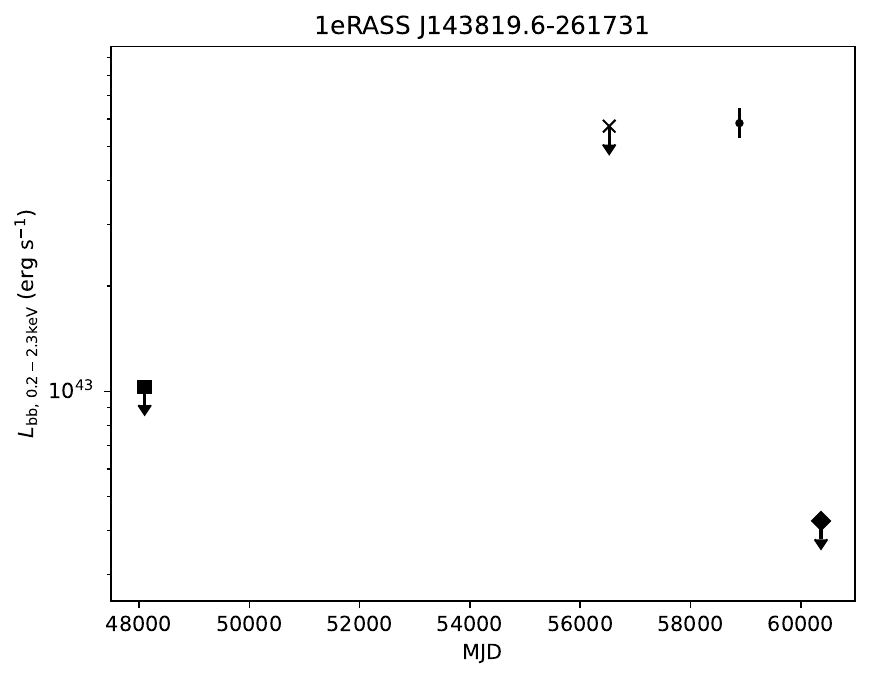}
\includegraphics[width=0.4\textwidth]{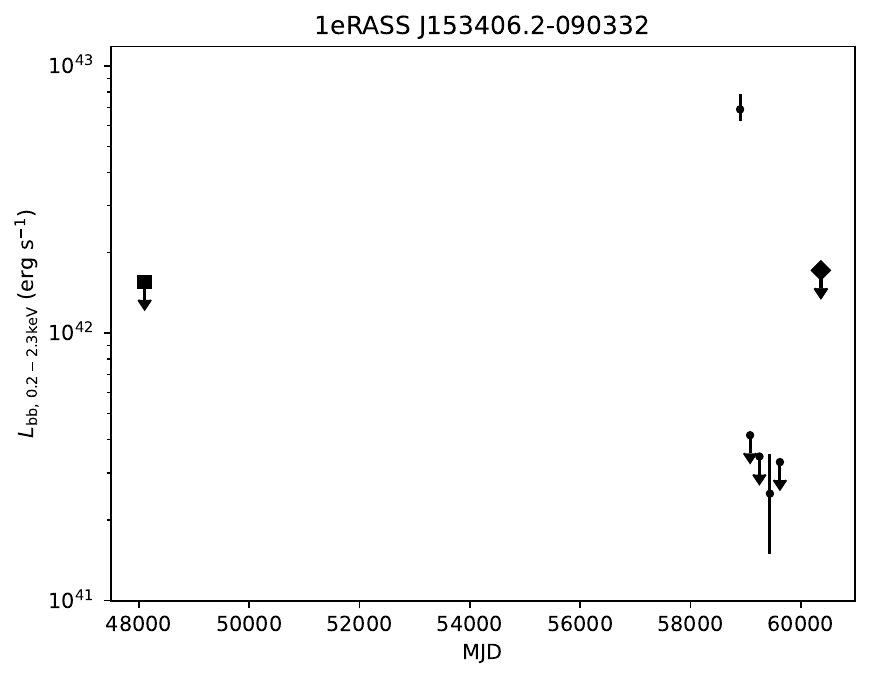}
\includegraphics[width=0.4\textwidth]{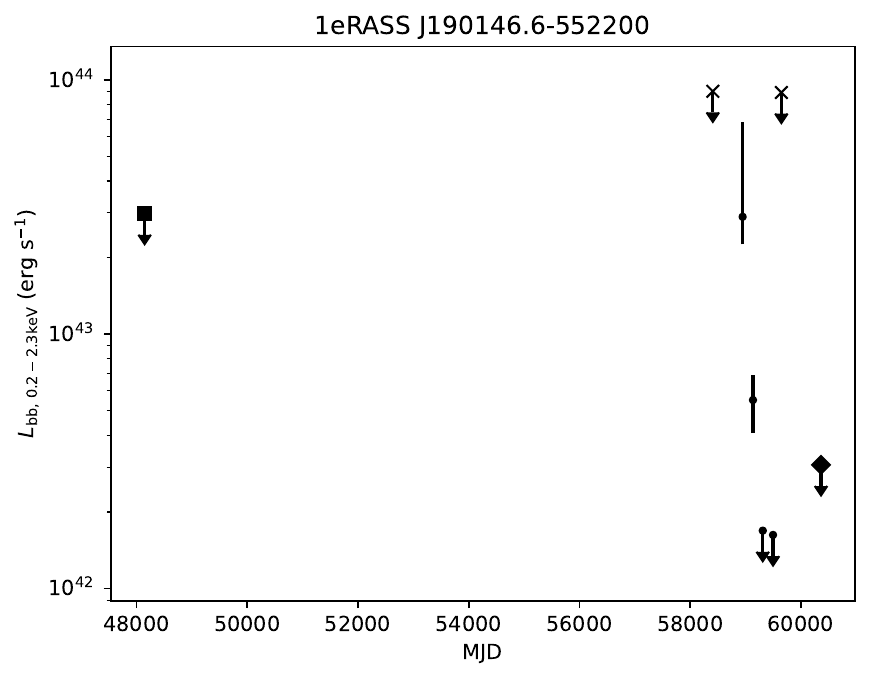}
\includegraphics[width=0.4\textwidth]{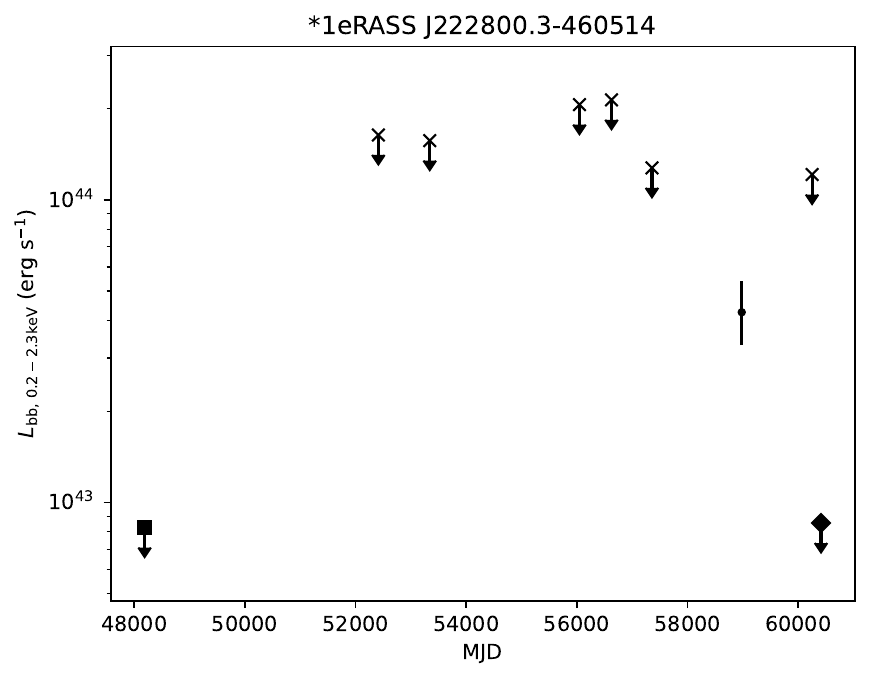}
\caption{The 0.2-2.3 keV X-ray luminosity light curves of our remaining candidate TDEs. Data from eROSITA, \textit{ROSAT}, \textit{XMM} and \textit{Swift} are marked with circles, squares, crosses and diamonds respectively. Note that we assume a redshift of $z = 0.1$  for 1eRASS J2228 (starred).}
\label{fig:xray_lcs}
\end{figure*}

A key feature of TDEs, and classically considered to be a smoking gun, is a decay in the fallback rate and the luminosity consistent with $t^{-5/3}$ \citep[e.g.][]{Rees88}. If the star has been only partially disrupted, the decay is expected to be somewhat quicker corresponding to $t^{-9/4}$ \citep{Nixon21}. This power law decay has been observed in several X-ray TDEs \citep[e.g. see Figure 1 of][]{Komossa04}.

Due to the fact that most of our sources are detected in only one or two epochs, we can constrain the decay index for only two of our sources, 1eRASS J0301 and 1eRASS J1436. We can directly fit these sources with a power law,
\begin{equation}
  L = L_0 ((t - t_p + t_{\rm fb}) / t_{\rm fb})^{\alpha},
  \label{eq:L_decay}
\end{equation}
where $L_0$ is the luminosity at the peak time, $t_p$ and $t_{\rm fb}$ is a fallback timescale. We fit only the post peak data fixing $\alpha$ to $-5/3$ using a least squares method. We repeat the fit assuming a partial disruption and therefore fix $\alpha=-9/4$. In the case of 1eRASS J0301, we also set the constraint $t_p > 58770$ i.e. peak is after the eROSITA-DE DR1 detection. These fits are plotted in Figure \ref{fig:xray_lcs_fitted}.

We find both sources are indeed compatible with both $t^{-5/3}$ and $t^{-9/4}$ decays. We also estimate the peak time of these sources, finding $t_p \sim {\rm MJD}~58600 - 58900$ for 1eRASS J0301 and $t_p \sim {\rm MJD} ~ 58780 - 58820$ for 1eRASS J1436. We note that due to the small number of datapoints and sensitivity to our intial guesses, these are only very weak constraints but they show that both these sources are broadly consistent with typical temporal behaviour from X-ray TDEs. Our inferred peak time is also earlier than that tabulated by \citet{Grotova25b}, however, theirs is the observational peak rather than the modelled peak as in our case.

\subsection{IR behaviour}
\label{sec:ir_behaviour}

The evidence of IR emission in our sample is surprising, particularly as it appears in four of our nine sources. This is significantly higher than the rate (3/26) found by \citet{Pasham25} for the optically selected ZTF TDE sample and a simple \textit{E}-test \citep{Krishnamoorthy04} suggests the probability that this work and theirs probe the same population of sources is $\sim0.2$. This could indicate that our method is biased towards a different population with crossover between both TDEs and AGN. The distant dusty torus surrounding an AGN could reprocess a much quicker nuclear flare to produce a slowly evolving IR light curve \citep{Yang19}. However, there is little other evidence for AGN activity in our hosts (see Section \ref{sec:host_properties}). An arguably more likely possibility, then, is that our supersoft X-ray selection criteria have a natural bias towards sources with IR counterparts.  With our small sample size, it is difficult to confirm this and the reason why this would be the case is unclear. We note that 10/31 of \citet{Grotova25b}'s sample have IR counterparts suggesting that X-ray bright TDEs may generally have a bias, particularly as these sources are not significantly softer than the rest of their sample. In this case, an \textit{E}-test indicates a probability $\sim0.65$ that the same population underlies both our samples while \citet{Langis25} also suggest a possible correlation between X-ray emission and infrared counterparts. We note that \citet{Pasham25} suggest that mid-infrared flares, similar to those we observe here, also indicate an elevated rate of QPEs. Together, this may suggest a link between supersoft X-ray spectra and QPEs and potentially an observing angle dependence for the latter.

\subsubsection{Dust shell modelling}

One other possible mechanism for the presence of IR is reprocessing in much closer dust surrounding the massive black hole being heated by emission from the region directly around the black hole \citep[e.g.][]{Lu16}. This can be simply modelled by assuming the dust is arranged in a spherical shell around the black hole. In spherical polar coordinates, a given dust grain will be positioned at radius and angle $R$, $\theta$. The infrared luminosity will therefore vary as \citep{Masterson24},
\begin{equation}
    L_{\rm IR} = \int_{{\rm min}(t - \tau,\, t_p)}^{t} \Psi(\tau) L_{\rm int}(t - \tau) d\tau ,
    \label{eq:Lir}
\end{equation}
where $\tau = (R_{\rm shell} / c) (1 - \cos \theta)$ is the delay between the intrinsic, $L_{\rm int}$, and IR, $L_{\rm IR}$, luminosity evolution, $\theta$ is the polar angle, and $\Psi(\tau)$ is uniform for $0 \leq \tau \leq 2R_{\rm shell}/c$ and 0 elsewhere. Because the rise time of the intrinsic luminosity is generally small compared to $\tau$, the IR luminosity will be dominated by post-peak emission and we therefore use Equation \ref{eq:L_decay} to model $L_{\rm int}$, fixing $\alpha = -5/3$. The resulting light curve has a rapid rise, starting from $\sim t_p$, and a slower decay.

We use the \textsc{emcee v3.1.4} Python package \citep{ForemanMackey13} to fit the \textit{W1} and \textit{W2} light curves simultaneously. Following \citet{Masterson24}, we do not model the temperature evolution directly but instead use a fixed ratio of \textit{W2} to \textit{W1} flux ($F_{\textit{W2}}/F_{\textit{W1}}$) inferred from blackbody fits to the light curve during the flare. While there is some temperature variation in the light curve, this fixed ratio is sufficient for our needs. We fit using 32 walkers across 5000 steps with the first 500 steps removed as burn-in. We also place an initial upper constraint of 1 pc on $R_{\rm shell}$ and assume a lower limit on $t_p$ equal to the last epoch where a constraining upper limit was obtained in both the \textit{W1} and \textit{W2} bands.

\begin{figure*}
    \centering
    \includegraphics[width=\columnwidth]{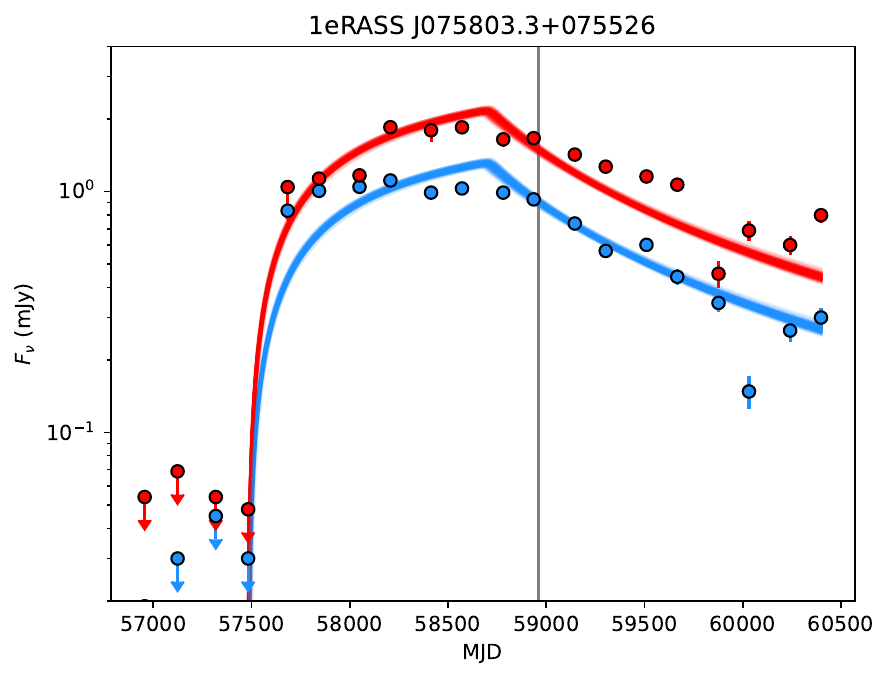}
    \includegraphics[width=\columnwidth]{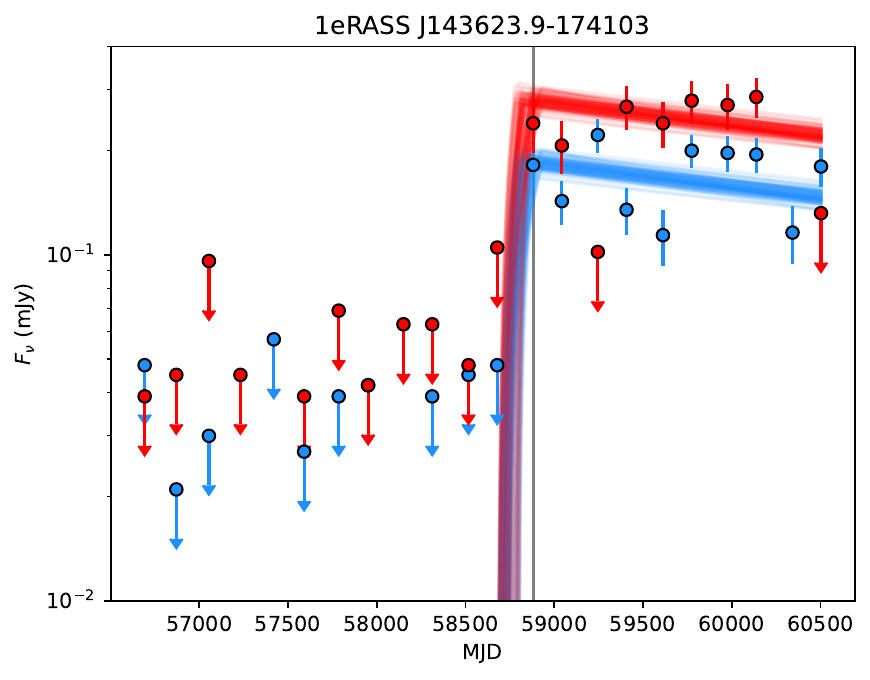}
    \includegraphics[width=\columnwidth]{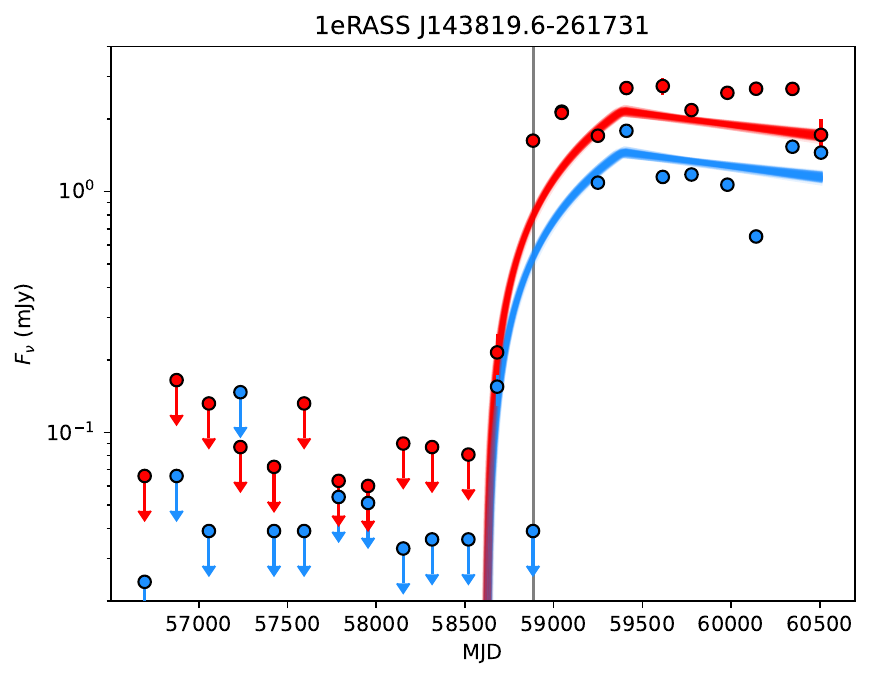}
    \includegraphics[width=\columnwidth]{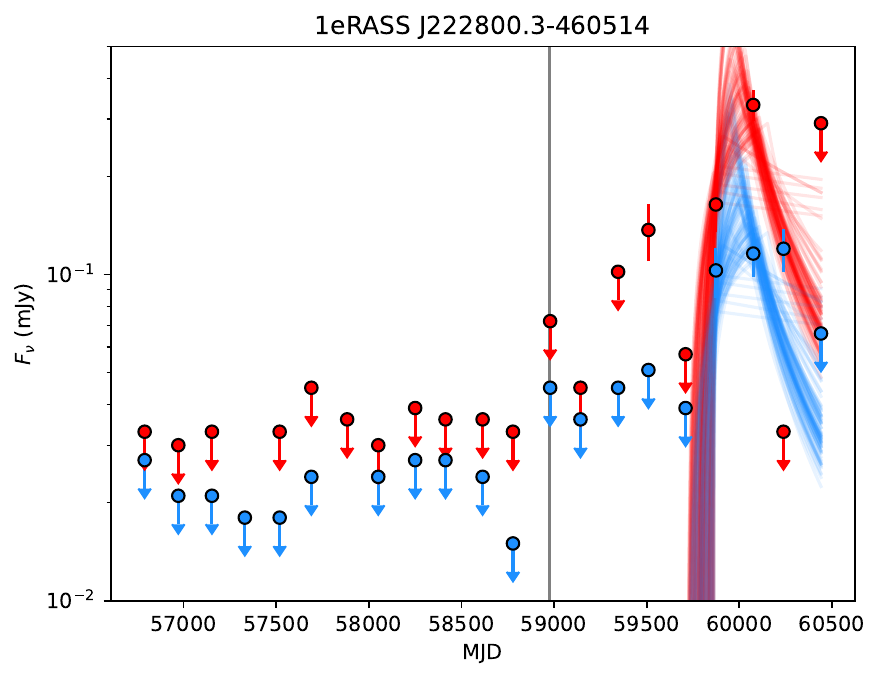}
    \includegraphics[width=3cm]{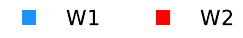}
    \caption{The infrared light curves of our four candidates with counterparts as observed by \textit{NEOWISE}. In all panels, the \textit{W1} and \textit{W2} bands are plotted in blue and red respectively, the grey vertical line indicates the time of the eROSITA-DE DR1 detections and upper limits are given to 3-$\sigma$. We also show 100 example light curves from our fitted spherical dust shell MCMC chain. In these fits, $R_{\rm shell}$ is fixed to $<1$ pc.}
    \label{fig:neowise_lcs}
\end{figure*}

\begin{table*}
    \caption{The results of our IR light curve fits. Note that $R_{\rm shell}$ is fixed to $<1$ pc and we have assumed $z=0.1$ for 1eRASS J2228 (and mark the radius and luminosity with a * accordingly).}
    \centering
    \renewcommand{\arraystretch}{1.4}
    \begin{tabular}{ccccccc}
    \hline
     TDE candidate & Estimated $T_{\rm BB}$ (K) & $F_{\textit{W2}}/F_{\textit{W1}}$ & $t_p$ (MJD) & $R_{\rm shell}$ (pc) & $\log(L_{\rm bb, pk})$ (erg s$^{-1}$) & $R_{\rm bb, pk}$ (pc) \\
    \hline
    1eRASS J0758 & 869 & 1.649 & $57482\pm3$ & $0.51\pm0.01$ & $43.13\pm0.01$ & $0.059\pm0.001$ \\
    1eRASS J1436 & 1013 & 1.509 & $58727\pm43$ & $0.05\pm0.02$ & $42.91\pm0.02$ & $0.034\pm0.001$ \\
    1eRASS J1438 & 901 & 1.485 & $58617\pm10$ & $0.32\pm0.01$ & $42.52\pm0.01$ & $0.027\pm0.001$ \\
    1eRASS J2228 & 727 & 2.162 & $59791\pm44$ & $0.10\pm0.04$ & $42.36^{+0.15}_{-0.09}$* & $0.035^{+0.007}_{-0.003}$* \\
    \hline
    \end{tabular}
    \label{tab:ir_fits}
\end{table*}

The fitted light curves are shown in Figure \ref{fig:neowise_lcs} and we summarise the inferred parameters in Table \ref{tab:ir_fits}. We also infer the radius from the peak luminosity in the modelled IR light curves, assuming the redshifts in Table \ref{tab:spectra} and that the IR SED is consistent with a blackbody, $R_{\rm bb} = (L_{\rm IR} / (4 \pi \sigma T^4))^{1/2}$, and the redshifts given in Table \ref{tab:spectra}. These are all significantly smaller than the $R_{\rm shell}$ inferred by our fits - a feature also noted by \citeauthor{Masterson24} and put down to the fact blackbodies are perfect emitters while the real efficiency is likely much lower.

We find, under our upper limit constraint, the values of $R_{\rm shell}$ for all our sources are broadly consistent with the distribution found by \citet{vanVelzen16}, \citet{Jiang21} and \citet{Masterson24}, although 1eRASS J0758 has a somewhat larger value compared to \citeauthor{Masterson24}'s sample. This is likely due to a combination of the selection effects both here and in that work, in particular, \citeauthor{Masterson24} note that their method is designed to select fast rising sources with accordingly smaller dust radii.

\begin{figure*}
    \centering
    \includegraphics[width=\columnwidth]{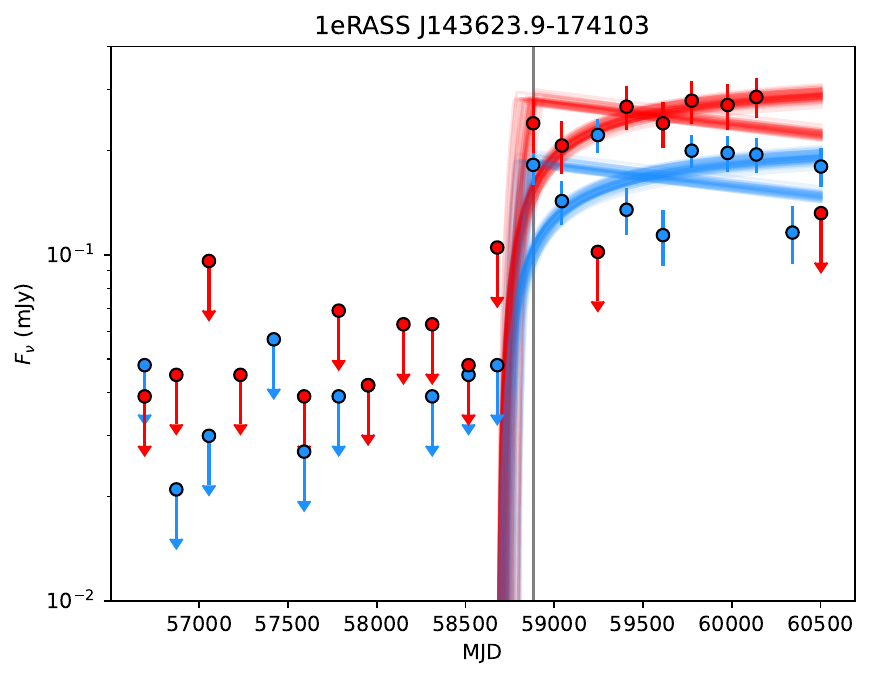}
    \includegraphics[width=\columnwidth]{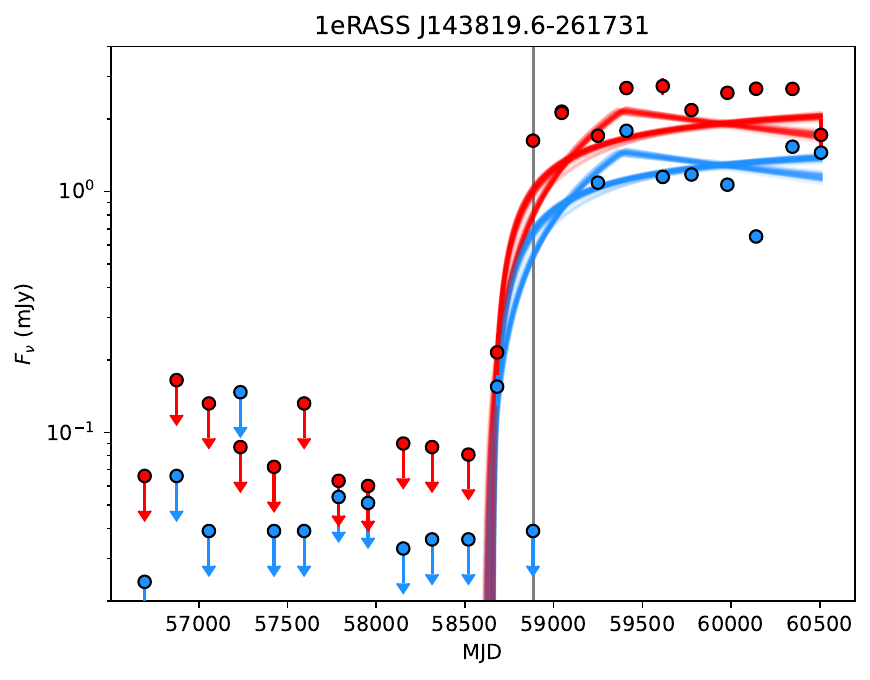}
    \includegraphics[width=3cm]{figs/legend.pdf}
    \caption{The infrared light curves of 1eRASS J1436 and 1eRASS J1438 as observed by \textit{NEOWISE}. In both panels,the \textit{W1} and \textit{W2} bands are plotted in blue and red respectively, the grey vertical line indicates the time of the eROSITA-DE DR1 detections and upper limits are given to 3-$\sigma$. We also show 100 example light curves from our fitted spherical dust shell MCMC chain. In these fits, $R_{\rm shell}$ is allowed to vary to $>1$ pc.}
    \label{fig:neowise_lcs_relaxed_R}
\end{figure*}

However, if the constraint on $R_{\rm shell}$ is relaxed, we find the fits to 1eRASS J1436 and 1eRASS J1438 also converge to significantly different values for $R_{\rm shell}$ of order $\gtrsim0.8$ pc. We show these light curves in Figure \ref{fig:neowise_lcs_relaxed_R}. While it is possible that the dust does indeed lie at a higher radius than typically observed in TDEs, another possibility is that these light curves feature a plateau similar to the QPE driven candidates of \citet{Pasham25} and the dust would therefore lie at much smaller radii consistent with our initial fits. We do  note, however, that 1eRASS J1436 and 1eRASS J1438 have much quicker rise times than the examples of AT 2019qiz and AT 2020ysg as shown in that work.

In the case of 1eRASS J1436, we also inferred $t_p$ from the X-ray light curve directly. However, the value found here implies the intrinsic luminosity peaks somewhat earlier than the X-ray luminosity. While this could well be a mathematical effect, it could also be physical. The model we assume is very simplistic and the radiative transfer calculations required for a full treatment are beyond the scope of this paper. It is also possible that the dust does start to be heated at much earlier times, for instance, while the TDE is still rising as our model assumes the effect of this phase to be negligible.

We do find that the fit to 1eRASS J0758 underpredicts the earliest detections of the flare, suggesting a quicker rise than allowed by our model. Allowing an earlier $t_p$ marginally improves the fit to those data but is inconsistent with the well constrained upper limits while the data show significant temperature evolution during the flare. This is therefore likely to be another case where the thin shell model cannot fully account for the observed behaviour, particularly the very steep initial rise and long plateau. A diffuse dust cloud or series of shells could help to explain these features, however, examining this further is beyond the scope of this paper.

The delay compared to the eROSITA detection of 1eRASS J2228 is also somewhat in contrast to typical expectations. However, we note that this source is close to the detection limit for our pipeline. A slower rise that begins at a significantly earlier time, but is not observable, is therefore plausible.

Our results are therefore broadly consistent with the rest of the IR TDE population with 1eRASS J1436 and 1eRASS J1438 representing possible outliers. Unfortunately, due to the lack of constraints in the optical data, we cannot infer the covering factor, $f_c$. This is an important parameter in determining the nuclear dust distribution in TDE host galaxies. Previous studies have shown $f_c$ to be significantly lower in TDE hosts than in the Milky Way or AGN \citep{vanVelzen16,Jiang21} which could imply a selection effect in the TDE sample and help explain why their host galaxies are quiescent.

\section{Host galaxy properties}
\label{sec:host_properties}

\begin{figure*}
\centering
\includegraphics[width=0.33\textwidth]{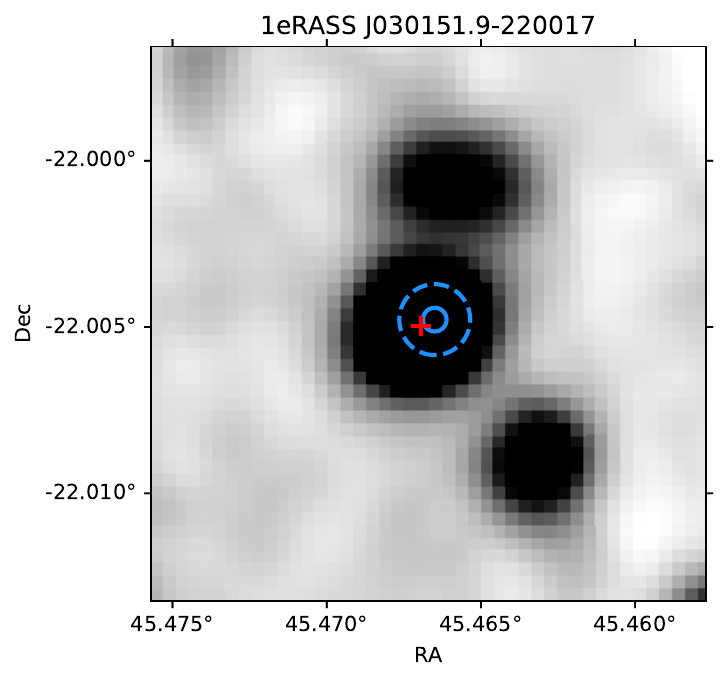}
\includegraphics[width=0.33\textwidth]{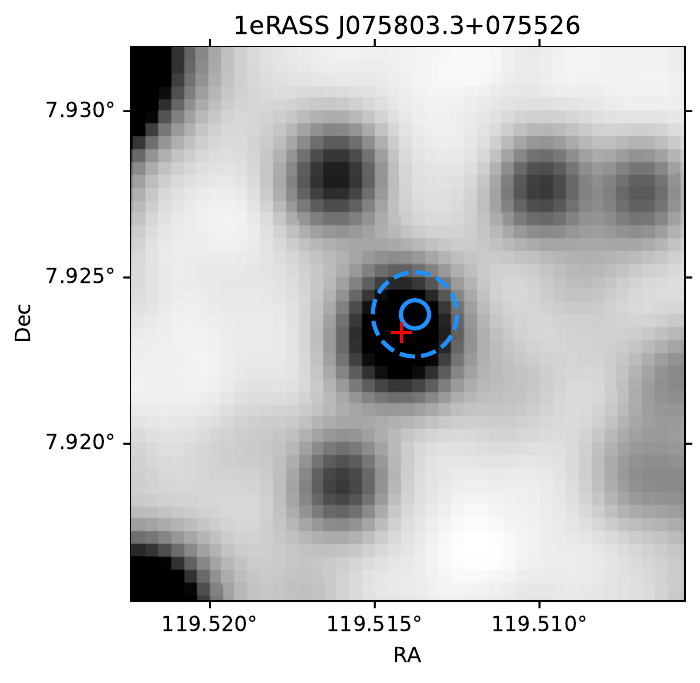}
\includegraphics[width=0.33\textwidth]{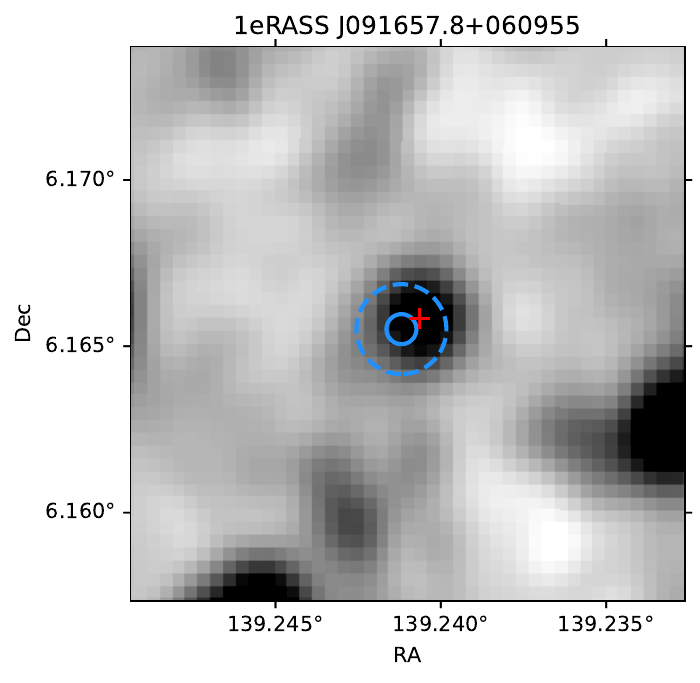}
\includegraphics[width=0.33\textwidth]{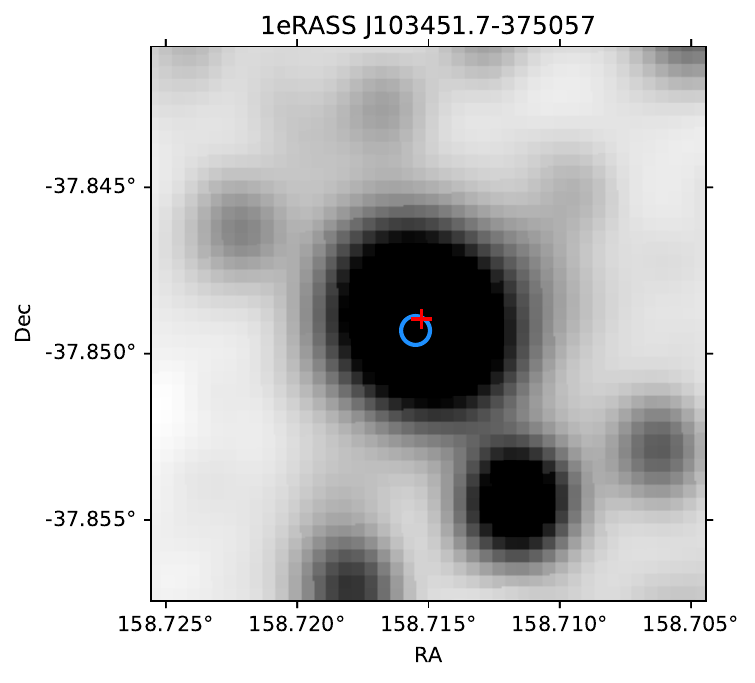}
\includegraphics[width=0.33\textwidth]{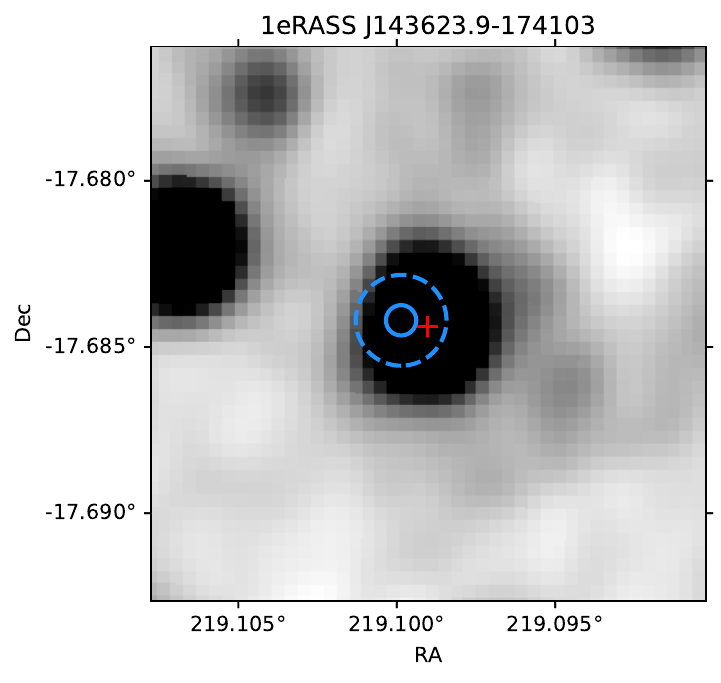}
\includegraphics[width=0.33\textwidth]{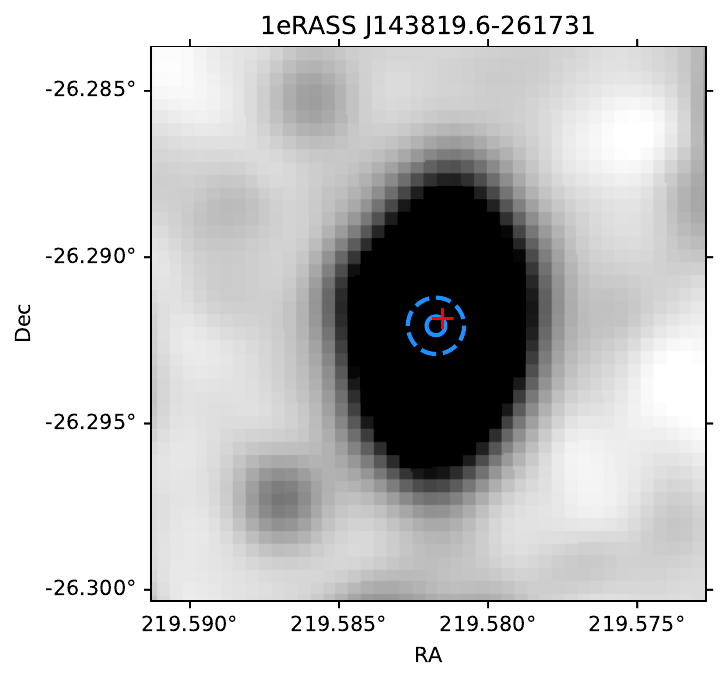}
\includegraphics[width=0.33\textwidth]{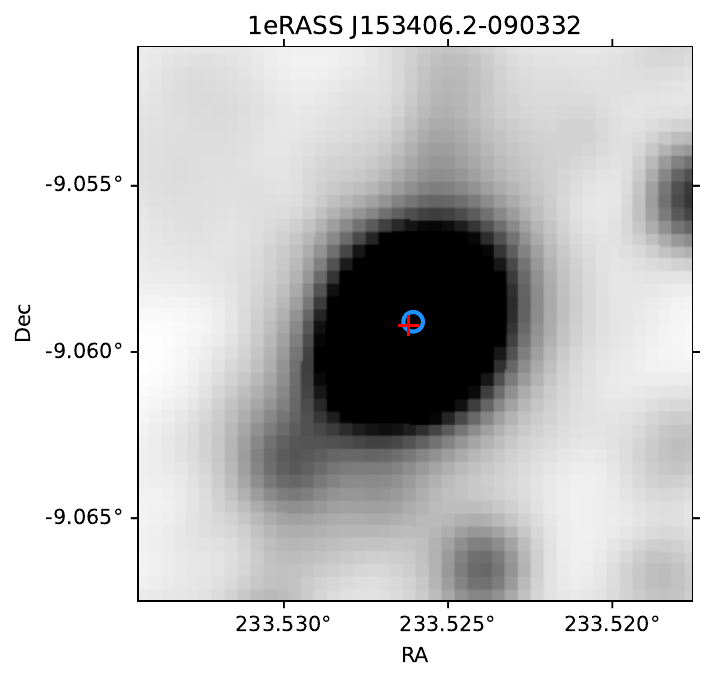}
\includegraphics[width=0.33\textwidth]{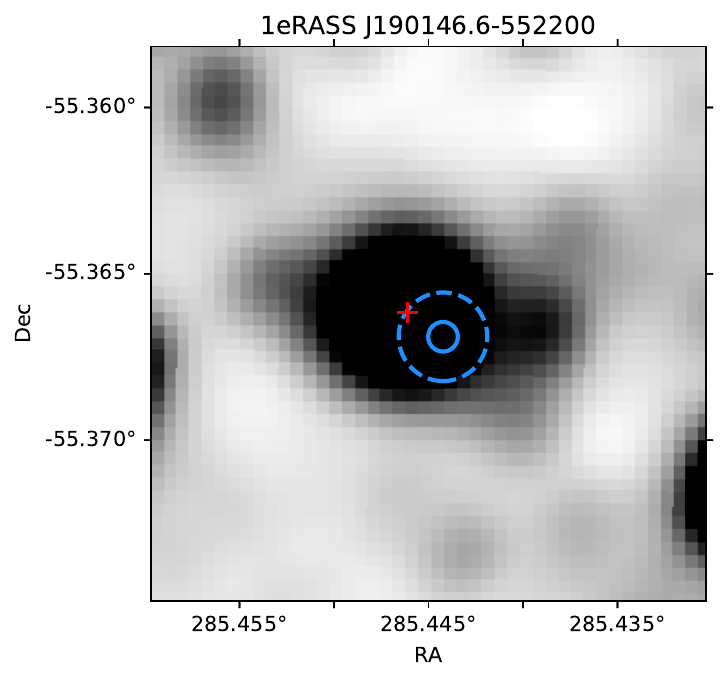}
\includegraphics[width=0.33\textwidth]{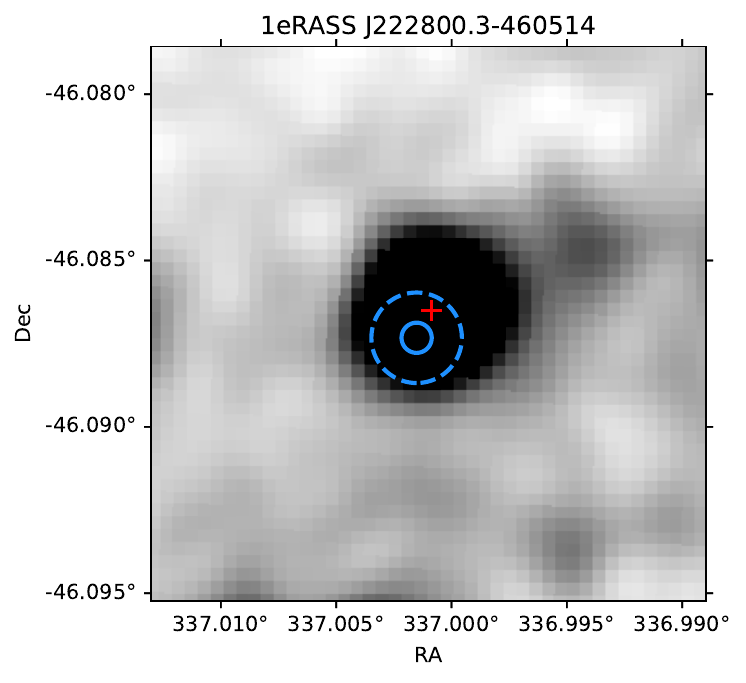}
\caption{\textit{WISE} \textit{W1} images of the putative host galaxies of our TDE candidates. Each panel is 1\arcmin~square. The blue circles indicate the 1-$\sigma$ (solid) and, where relevant, 3-$\sigma$ (dashed) eROSITA position errors. In all cases, the eROSITA positions are consistent to 3-$\sigma$ or less with the \textit{W1} centroid (red cross).}
\label{fig:wise_plots}
\end{figure*}

\begin{figure}
    \centering
    \includegraphics[width=\columnwidth]{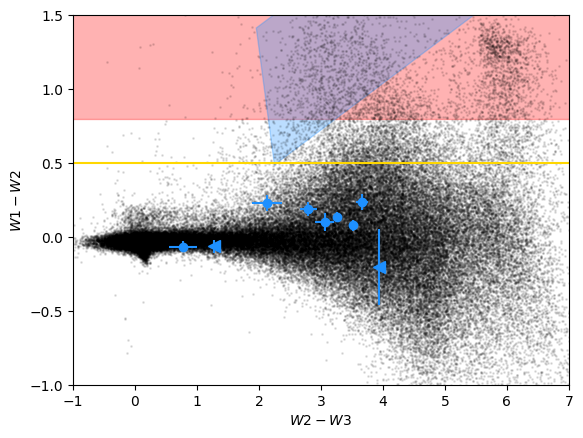}
    \caption{The \textit{WISE} colours of our candidates' putative hosts (blue) compared to a random sample of sources from the AllWISE catalogue (black). The AGN criteria of \citet{Stern12} and \citet{Mateos12} are shaded in red and blue respectively and the criterion of \citet{Mingo16} and \citet{Mingo19} is indicated with the gold line. Note that sources marked with triangles are not detected in \textit{W3} and therefore their x-coordinates are upper limits.}
    \label{fig:wise_colours}
\end{figure}

Finally, we analyse the photometric properties of our host galaxies. All of our candidates are spatially consistent with \textit{WISE} sources, detected in \textit{W1} and \textit{W2}, and we show \textit{W1} cutouts of the hosts relative to the eROSITA source position in Figure \ref{fig:wise_plots}. Most of those sources are also detected in \textit{W3} and the remaining source has a reasonably constraining \textit{W3} upper limit available instead. Comparing the colours across these bands allows us to roughly classify the hosts and investigate the possibility of AGN activity \citep{Wright10}. We plot our candidates' host colours compared to a random sample of sources from the AllWISE catalogue \citep{Cutri13} and the AGN criteria of \citet{Stern12} and \citet{Mateos12} in Figure \ref{fig:wise_colours}. We find all of our host candidates are consistent with classification as spiral galaxies, typical for the observed TDE population \citep[see e.g. Figure 2 of][]{Hinkle24}. The hosts of 1eRASS J1534 and 1eRASS J1438 are also consistent with elliptical galaxies and the hosts of 1eRASS J0916, 1eRASS J1034, 1eRASS J1901 and 1eRASS J2228 are also consistent with luminous infrared galaxies (LIRGs), some of which have been shown to host AGN and GRBs \citep[e.g.][]{EylesFerris23}. In addition, 1eRASS J1034's host, LEDA 3081763, is broadly consistent with starburst galaxies, consistent with its classification as an emission line galaxy \citep{Chen22}, although we note this could also indicate the presence of an AGN.

We identified further photometry in SkyMapper Data Release 4 \citep{Onken24}, with detections varying across the SkyMapper \textit{u}, \textit{v}, \textit{g}, \textit{r}, \textit{i} and \textit{z} filters to construct our host galaxies' spectral energy distributions (SEDs). We correct the \textit{WISE} and SkyMapper photometry for Galactic extinction taken from the maps of \citet{Schlafly11} using the \texttt{dust\_extinction v1.5} package \citep{Gordon24b} and the \citet{Gordon23} Milky Way model with $R_V = 3.1$ \citep[see also][]{Gordon09,Fitzpatrick19,Gordon21,Decleir22}, and fit the host photometry using the \textsc{CIGALE} code \citep{Boquien19,Yang20,Yang22}. We use a model consisting of a double exponential star formation history with stellar and dust emission from \citet{Bruzual03} and \citet{Dale14}. Dust attenuation is applied using the model of \citet{Calzetti00} and we use the redshifts in Table \ref{tab:spectra}, again assuming $z=0.1$ for 1eRASS J2228. We exclude the \textit{Swift}/UVOT data from these fits to test for for lingering TDE emission, for instance from the late time UV plateaus commonly found among the TDE population \citep[e.g.][]{Mummery24a,Mummery24b}.

\begin{figure}
    \centering
    \includegraphics[width=\columnwidth]{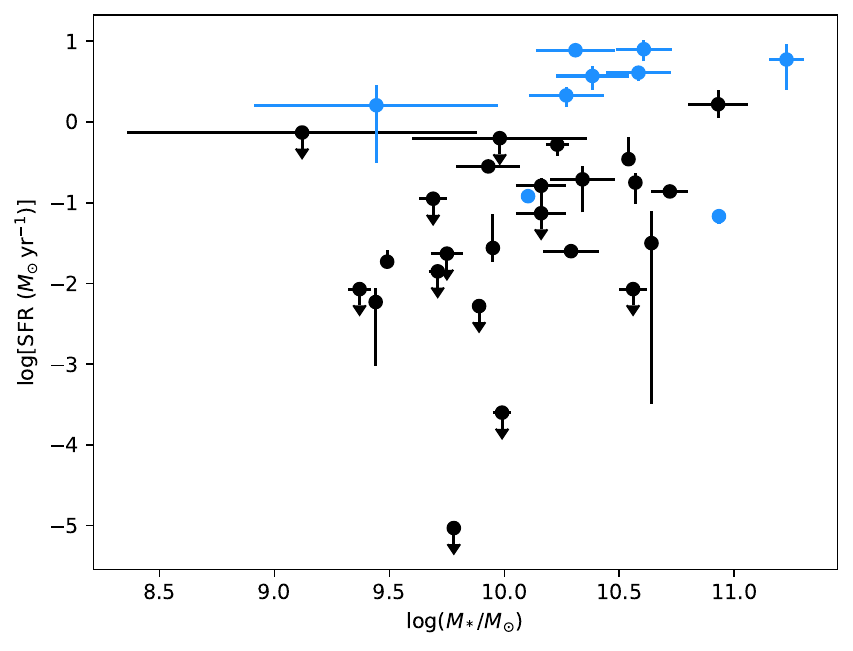}
    \caption{The stellar masses and star formation rates of our candidates putative host galaxies (blue) compared to the sample of \citet{Hinkle21} (black).}
    \label{fig:hinkle_comp}
\end{figure}

We find that no additional AGN contribution was required to produce a good fit to our assembled photometry, implying that our host candidates are indeed inactive galaxies. These fits infer stellar masses and star formation rates (SFRs) and, using the relation of \citet{Greene20}:
\begin{equation}
    \log(M_{\bullet} / M_{\odot}) = 7.35 \pm 0.22 + (1.61 \pm 0.24)\log(M_* / M_O)\,,
    \label{eq:M_*_to_M_bh}
\end{equation} 
where $M_O = 3\times10^{10} M_{\odot}$,
we also infer masses of the central black hole. This relation has been shown to best match the properties of other TDE hosts \citep{Hammerstein23b}. We present the results of these fits in Table \ref{tab:hosts} and compare to the results of \citet{Hinkle21} in Figure \ref{fig:hinkle_comp}. We find that the \textit{Swift}/UVOT photometry is also broadly consistent with extrapolations from our fits to the higher wavelength data indicating any late time UV emission from the TDE is negligible compared to the host galaxy.

\begin{table*}
    \caption{The derived properties of our candidates' putative hosts. Note that we assumed the redshift of 1eRASS J2228 to be $z=0.1$.}
    \centering
    \renewcommand{\arraystretch}{1.4}
    \begin{tabular}{cccccc}
    \hline
     TDE candidate & z & $A_{V,\,{\rm Gal}}$ (mag) & $\log (M_* / M_{\odot})$ & $\log (M_{\bullet} / M_{\odot})$ & $\log [{\rm SFR} (M_{\odot} {\rm yr}^{-1})]$\\
    \hline
1eRASS J0301 & 0.0792$^1$ & 0.068 & $10.27^{+0.12}_{-0.16}$ & $7.02^{+0.43}_{-0.57}$ & $0.33^{+0.11}_{-0.15}$ \\
1eRASS J0758 & 0.0955$^2$ & 0.052 & $10.38^{+0.12}_{-0.16}$ & $7.20^{+0.40}_{-0.53}$ & $0.57^{+0.12}_{-0.17}$ \\
1eRASS J0916 & 0.091$^3$ & 0.111 & $9.44^{+0.23}_{-0.53}$ & $5.69^{+0.79}_{-1.45}$ & $0.21^{+0.26}_{-0.71}$ \\
1eRASS J1034 & 0.06291$^4$ & 0.223 & $10.61^{+0.10}_{-0.12}$ & $7.56^{+0.32}_{-0.42}$ & $0.90^{+0.11}_{-0.15}$ \\
1eRASS J1436 & 0.1925$^3$ & 0.232 & $11.23^{+0.06}_{-0.08}$ & $8.56^{+0.13}_{-0.18}$ & $0.77^{+0.20}_{-0.37}$ \\
1eRASS J1438 & 0.04732$^4$ & 0.254 & $10.93\pm0.02$ & $8.09^{+0.14}_{-0.15}$ & $-1.17^{+0.08}_{-0.10}$ \\
1eRASS J1534 & 0.02404$^4$ & 0.354 & $10.10\pm0.02$ & $6.75^{+0.34}_{-0.35}$ & $-0.92\pm0.04$ \\
1eRASS J1901 & 0.059$^3$ & 0.289 & $10.58^{+0.11}_{-0.14}$ & $7.52^{+0.34}_{-0.46}$ & $0.61^{+0.09}_{-0.11}$ \\
1eRASS J2228 & --- & 0.035 & $10.31^{+0.12}_{-0.17}$ & $7.08^{+0.43}_{-0.58}$ & $0.89^{+0.06}_{-0.07}$ \\
    \hline
    \multicolumn{6}{l}{$^1$\citet{Gaia23}, $^2$\citet{Alam15}, $^3$\citet{Grotova25}, $^4$\citet{Jones09}.}\\
    \end{tabular}
    \label{tab:hosts}
\end{table*}

For sources in common with \citet{Grotova25b}, our results are consistent and the stellar masses of our complete sample are consistent with those identified by both \citet{Hammerstein21} and \citet{Hinkle21}. However, the black hole masses are somewhat higher than typical for TDEs. This suggests that Equation \ref{eq:M_*_to_M_bh} may not accurately reflect the entire population of TDE hosts, particularly X-ray detected examples, as \citet{Hammerstein23b}'s sample primarily consists of optically selected TDEs. The SFRs are also somewhat higher than those derived by \citet{Hinkle21}. The discrepancy could reinforce the suggestion that X-ray and optical TDE hosts differ in some ways but could also be an effect of the different SED fitting methods. We also note that the SFRs we infer do appear broadly consistent with the general stellar mass-SFR trend shown by the galaxies in \citet{Hinkle21}'s sample. 1eRASS J1034's host has a slightly higher SFR than most of our sample, again reflecting its likely nature as a galaxy with emission lines driven by star formation. 1eRASS J2228's host also has an elevated SFR, however its redshift has been assumed and therefore the inferred SFR is less reliable.

TDE hosts are often found to be consistent with post-starburst green valley galaxies \citep[e.g.][]{Hammerstein21}. Typically, these are identified through their mass and \textit{u} - \textit{g} colour in the SDSS photometric system. Unfortunately, while SkyMapper's \textit{u} and \textit{v} filters cover a suitable part of the spectrum, none of our sources' hosts were detected in these filters. However, the hosts of 1eRASS J0758 and 1eRASS J0916 are within the SDSS footprint. We find that 1eRASS J0916's host is indeed compatible with a green valley classification while 1eRASS J0758's host is more consistent with red sequence galaxies.

Despite this, our results are consistent with the picture that these galaxies are generally spirals with properties broadly aligning with the TDE host population.

\section{Conclusions}
\label{sec:conclusions}

In this work, we have shown that TDE candidates in catalogued X-ray data can be identified through their supersoft X-ray spectra and we present nine such candidates in eROSITA-DE DR1. Four of these are present in the \eroextra~catalogue of eROSITA nuclear transients but five are entirely new X-ray TDE candidates. We summarise our conclusions below:
\begin{itemize}
    \item The X-ray properties of our candidates are broadly consistent with the wider population of X-ray TDEs, displaying soft thermal spectra and consistent with $t^{-5/3}$ decays where there are sufficient data to evaluate this. We also present additional \textit{Swift} observations which show our candidates to have faded significantly and constrain the spectral evolution of one source, 1eRASS J0301, finding tentative evidence of hardening.
    \item Two of our candidates have optical counterparts, however, gaps in coverage means we cannot investigate the optical properties of these counterparts further.
    \item Remarkably, four of our sources have IR counterparts detected by \textit{NEOWISE}, hinting that our supersoft X-ray selection criteria have a natural bias towards such sources. We use a simple dust shell model to fit the IR light curves to constrain the environments surrounding these sources. Our modelling indicates dust shells with radii of $\sim$a few 0.1 pc for all sources, consistent with the sample studied by \citet{Masterson24}. However, two sources are also consistent with dust shells at much greater radii of $\gtrsim0.8$ pc.
    \item The properties of our candidates' host galaxies are broadly consistent with the general population of TDE hosts.
\end{itemize}

While there is crossover with sources identified by \eroextra~and \citet{Grotova25b}, we demonstrate that spectral selection can also be used to discover TDEs and similar methods could be used to identify potential TDEs on the fly in the future. We eagerly await the release of further eROSITA-DE catalogues to refine our selection criteria and continue expanding the TDE sample.

%%%%%%%%%%%%%%%%%%%%%%%%%%%%%%%%%%%%%%%%%%%%%%%%%%
\section*{Acknowledgements}

We thank the anonymous referee for their constructive comments and DJ Pasham for useful discussion.

RLCS acknowledges support from Leverhulme Trust grant RPG-2023-240. PTO acknowledges support from UKRI under grant ST/W000857/1. KLP and PAE acknowledge funding from the UK Space Agency.

This work is based on data from eROSITA, the soft X-ray instrument aboard SRG, a joint Russian-German science mission supported by the Russian Space Agency (Roskosmos), in the interests of the Russian Academy of Sciences represented by its Space Research Institute (IKI), and the Deutsches Zentrum für Luft- und Raumfahrt (DLR). The SRG spacecraft was built by Lavochkin Association (NPOL) and its subcontractors, and is operated by NPOL with support from the Max Planck Institute for Extraterrestrial Physics (MPE). The development and construction of the eROSITA X-ray instrument was led by MPE, with contributions from the Dr. Karl Remeis Observatory Bamberg \& ECAP (FAU Erlangen-Nuernberg), the University of Hamburg Observatory, the Leibniz Institute for Astrophysics Potsdam (AIP), and the Institute for Astronomy and Astrophysics of the University of Tübingen, with the support of DLR and the Max Planck Society. The Argelander Institute for Astronomy of the University of Bonn and the Ludwig Maximilians Universität Munich also participated in the science preparation for eROSITA.

This work made use of data supplied by the UK \textit{Swift} Science Data Centre at the University of Leicester; the VizieR catalogue access tool, CDS, Strasbourg, France (DOI : 10.26093/cds/vizier); the Wide-field Infrared Survey Explorer, which is a joint project of the University of California, Los Angeles, and the Jet Propulsion Laboratory/California Institute of Technology, funded by the National Aeronautics and Space Administration; the Asteroid Terrestrial-impact Last Alert System (ATLAS) project; the ZTF forced-photometry service; and SkyMapper.

The ATLAS project is primarily funded to search for near earth asteroids through NASA grants NN12AR55G, 80NSSC18K0284, and 80NSSC18K1575; byproducts of the NEO search include images and catalogs from the survey area. This work was partially funded by Kepler/K2 grant J1944/80NSSC19K0112 and HST GO-15889, and STFC grants ST/T000198/1 and ST/S006109/1. The ATLAS science products have been made possible through the contributions of the University of Hawaii Institute for Astronomy, the Queen’s University Belfast, the Space Telescope Science Institute, the South African Astronomical Observatory, and The Millennium Institute of Astrophysics (MAS), Chile.

The ZTF forced-photometry service was funded under the Heising-Simons Foundation grant \#12540303 (PI: Graham).

The national facility capability for SkyMapper has been funded through ARC LIEF grant LE130100104 from the Australian Research Council, awarded to the University of Sydney, the Australian National University, Swinburne University of Technology, the University of Queensland, the University of Wwd = 'WDJ041102.17-035822.59'estern Australia, the University of Melbourne, Curtin University of Technology, Monash University and the Australian Astronomical Observatory. SkyMapper is owned and operated by The Australian National University's Research School of Astronomy and Astrophysics. The survey data were processed and provided by the SkyMapper Team at ANU. The SkyMapper node of the All-Sky Virtual Observatory (ASVO) is hosted at the National Computational Infrastructure (NCI). Development and support of the SkyMapper node of the ASVO has been funded in part by Astronomy Australia Limited (AAL) and the Australian Government through the Commonwealth's Education Investment Fund (EIF) and National Collaborative Research Infrastructure Strategy (NCRIS), particularly the National eResearch Collaboration Tools and Resources (NeCTAR) and the Australian National Data Service Projects (ANDS).

For the purpose of open access, the author has applied a Creative Commons Attribution (CC BY) licence to the Author Accepted Manuscript version arising from this submission.

\section*{Data Availability}

All data used in this work are available publicly from their respective repositories (see footnotes in text for urls).



\bibliographystyle{mnras}
\bibliography{eROSITA_tde}

%%%%%%%%%%%%%%%%%%%%%%%%%%%%%%%%%%%%%%%%%%%%%%%%%%

%%%%%%%%%%%%%%%%% APPENDICES %%%%%%%%%%%%%%%%%%%%%

\appendix

\section{X-ray spectral fitting results}
\label{sec:Xray_appendix}

\begin{table*}
  \caption{The results of our spectral fitting. $N_{\rm H}$ is given in units of $10^{20}$ cm$^{-2}$ and $C/\nu$ is the Cash-statistic over degrees of freedom. We bold $C/\nu$ to indicate the statistically favoured model, i.e. where $C/\nu$ is closest to 1, based on the non-rounded value. Luminosities are calculated using the blackbody model and the redshifts in Table \ref{tab:sample} or for values marked with *s, we assume a redshift of $z=0.1$. Errors are given to 90\% confidence.}
  \label{tab:spectra}
  \centering
  \renewcommand{\arraystretch}{1.4}
  \begin{tabular}{cccccccccc}
    \hline
    MJD & Mission & & \multicolumn{3}{c}{Blackbody} & \multicolumn{3}{c}{Power law}  & $\log\left(\frac{L_{\rm bb,\,0.2-2.3~keV}}{{{\rm erg~s}^{-1}}}\right)$ \\
& & $N_{\rm H, Gal}$ & $N_{\rm H, host}$ & kT (eV) & $C/\nu$ & $N_{\rm H, Host}$ & $\Gamma$ & $C/\nu$ & \\
    \hline
\multicolumn{10}{c}{1eRASS J0301}\\
48257.0 & \textit{ROSAT}/PSPC &&&&&&&& $<42.86$ \\
54665.2 & \textit{XMM} &&&&&&&& $<43.76$ \\
55052.4 & \textit{XMM} &&&&&&&& $<43.93$ \\
57611.9 & \textit{XMM} &&&&&&&& $<44.04$ \\
58870.3 & \textit{SRG}/eROSITA & 1.9 & $<1.6$ & $101\pm9$ & \textbf{75/88} & $6.8^{+3.8}_{-3.3}$ & $5.0^{+0.8}_{-0.7}$ & 60/88 & $43.33^{+0.10}_{-0.06}$\\
59032.8 & \textit{XMM} &&&&&&&& $44.10^{+0.13}_{-0.19}$ \\
60364.4 & \textit{Swift}/XRT &&&&&&&& $42.72^{+0.13}_{-0.20}$ \\
60645.1 & \textit{Swift}/XRT & 1.9 & $<0.4$ & $191^{+39}_{-30} $ & 56/35 & $<19.0$ & $3.2^{+1.2}_{-0.5}$ & \textbf{41/35} & $42.70^{+0.13}_{-0.12}$ \\
\hline
\multicolumn{10}{c}{1eRASS J0758}\\
48171.0 & \textit{ROSAT}/PSPC &&&&&&&& $<43.32$ \\
55845.4 & \textit{XMM} &&&&&&&& $<43.90$ \\
58962.7 & \textit{SRG}/eROSITA & 2.3 & $<1.5$ & $162^{+25}_{-21}$ & \textbf{45/63} & $2.1^{+5.7}_{-2.1}$ & $3.2^{+0.8}_{-0.6}$ & 33/63 & $43.39\pm0.09$\\
60703.4 & \textit{Swift}/XRT &&&&&&&& $<43.19$ \\
\hline
\multicolumn{10}{c}{1eRASS J0916}\\
48185.0 & \textit{ROSAT}/PSPC &&&&&&&& $<42.98$ \\
48936.2 & \textit{ROSAT}/PSPC &&&&&&&& $<42.35$ \\
52398.6 & \textit{XMM} &&&&&&&& $<44.56$ \\
55508.9 & \textit{XMM} &&&&&&&& $<44.42$ \\
58960.8 & \textit{XMM} &&&&&&&& $<46.70$ \\
58977.4 & \textit{SRG}/eROSITA & 3.5 & $<2.0$ & $65^{+10}_{-9}$ & \textbf{32/34} & $2.8^{+5.6}_{-2.8}$ & $6.1^{+1.9}_{-1.3}$ & 28/34 & $43.66^{+0.21}_{-0.14}$\\
59162.1 & \textit{SRG}/eROSITA &&&&&&&& $<42.76$ \\
59345.4 & \textit{SRG}/eROSITA &&&&&&&& $<42.92$ \\
59529.1 & \textit{SRG}/eROSITA &&&&&&&& $42.44^{+0.19}_{-0.35}$ \\
60367.0 & \textit{Swift}/XRT &&&&&&&& $<42.91$ \\
\hline
\multicolumn{10}{c}{1eRASS J1034}\\
48083.0 & \textit{ROSAT}/PSPC &&&&&&&& $<42.92$ \\
56645.2 & \textit{XMM} &&&&&&&& $<43.78$ \\
57354.8 & \textit{XMM} &&&&&&&& $<44.06$ \\
57393.7 & \textit{XMM} &&&&&&&& $<43.79$ \\
59007.8 & \textit{SRG}/eROSITA & 6.4 & $<2.1$ & $121^{+16}_{-12}$ & \textbf{43/61} & $0.2^{+7.0}_{-0.2}$ & $4.0^{+0.9}_{-0.4}$ & 39/61 & $43.02\pm0.11$\\
59008.9 & \textit{XMM} &&&&&&&& $<44.26$ \\
59369.9 & \textit{XMM} &&&&&&&& $<44.03$ \\
60704.2 & \textit{Swift}/XRT &&&&&&&& $<42.52$ \\
\hline
  \end{tabular}
 \end{table*}
 
\begin{table*}
  \contcaption{}
  \centering
  \renewcommand{\arraystretch}{1.4}
  \begin{tabular}{cccccccccc}
    \hline
    MJD & Mission & $N_{\rm H, Gal}$ & \multicolumn{3}{c}{Blackbody} & \multicolumn{3}{c}{Power law}  & $\log\left(\frac{L_{\rm bb,\,0.2-2.3~keV}}{{{\rm erg~s}^{-1}}}\right)$ \\
& & & $N_{\rm H, host}$ & kT (eV) & $C/\nu$ & $N_{\rm H, Host}$ & $\Gamma$ & $C/\nu$ \\
    \hline
\multicolumn{10}{c}{1eRASS J1436}\\
48102.0 & \textit{ROSAT}/PSPC &&&&&&&& $<44.04$ \\
58881.1 & \textit{SRG}/eROSITA & 8.2 & $<6.8$ & $111\pm14$ & 34/53 & $11.4^{+19.1}_{-11.4}$ & $5.6^{+1.7}_{-1.2}$ & \textbf{34/53} & $44.18^{+0.25}_{-0.11}$\\
59064.7 & \textit{SRG}/eROSITA &&&&&&&& $43.20^{+0.14}_{-0.22}$ \\
59238.8 & \textit{SRG}/eROSITA &&&&&&&& $43.06^{+0.19}_{-0.34}$ \\
59427.6 & \textit{SRG}/eROSITA &&&&&&&& $<43.55$ \\
59609.6 & \textit{SRG}/eROSITA &&&&&&&& $<43.55$ \\
60368.2 & \textit{Swift}/XRT &&&&&&&& $<43.97$ \\
\hline
\multicolumn{10}{c}{1eRASS J1438}\\
48102.0 & \textit{ROSAT}/PSPC &&&&&&&& $<43.01$ \\
56524.2 & \textit{XMM} &&&&&&&& $<43.76$ \\
58885.8 & \textit{SRG}/eROSITA & 8.9 & 0.00 & $99\pm4$ & 216/149 & $3.2^{+2.7}_{-2.4}$ & $5.0\pm0.4$ & \textbf{159/149} & $43.77\pm0.04$\\
60365.0 & \textit{Swift}/XRT &&&&&&&& $<42.63$ \\
\hline
\multicolumn{10}{c}{1eRASS J1534}\\
48102.0 & \textit{ROSAT}/PSPC &&&&&&&& $<42.19$ \\
58900.2 & \textit{SRG}/eROSITA & 12.5 & $<1.1$ & $120\pm6$ & 164/150 & $7.1^{+5.6}_{-4.5}$ & $4.9\pm0.5$ & \textbf{143/150} & $42.84^{+0.06}_{-0.05}$\\
59081.0 & \textit{SRG}/eROSITA &&&&&&&& $<41.62$ \\
59250.5 & \textit{SRG}/eROSITA &&&&&&&& $<41.54$ \\
59440.4 & \textit{SRG}/eROSITA &&&&&&&& $41.40^{+0.15}_{-0.23}$ \\
59622.1 & \textit{SRG}/eROSITA &&&&&&&& $<41.52$ \\
60362.9 & \textit{Swift}/XRT &&&&&&&& $<42.24$ \\
60367.1 & \textit{Swift}/XRT &&&&&&&& $<42.24$ \\
\hline
\multicolumn{10}{c}{1eRASS J1901}\\
48146.0 & \textit{ROSAT}/PSPC &&&&&&&& $<43.47$ \\
58410.2 & \textit{XMM} &&&&&&&& $<43.95$ \\
58948.6 & \textit{SRG}/eROSITA & 8.2 & $<7.4$ & $88^{+11}_{-14}$ & \textbf{48/56} & $9.4^{+13.2}_{-8.8}$ & $5.9^{+1.9}_{-1.5}$ & 48/56 & $43.46^{+0.37}_{-0.11}$\\
59135.5 & \textit{SRG}/eROSITA &&&&&&&& $42.74^{+0.10}_{-0.13}$ \\
59312.3 & \textit{SRG}/eROSITA &&&&&&&& $<42.23$ \\
59497.6 & \textit{SRG}/eROSITA &&&&&&&& $<42.21$ \\
59648.3 & \textit{XMM} &&&&&&&& $<43.95$ \\
60364.5 & \textit{Swift}/XRT &&&&&&&& $<42.49$ \\
\hline
\multicolumn{10}{c}{1eRASS J2228}\\
48187.0 & \textit{ROSAT}/PSPC &&&&&&&& *$<42.92$ \\
52410.6 & \textit{XMM} &&&&&&&& *$<44.21$ \\
53338.9 & \textit{XMM} &&&&&&&& *$<44.20$ \\
56045.0 & \textit{XMM} &&&&&&&& *$<44.31$ \\
56623.9 & \textit{XMM} &&&&&&&& *$<44.33$ \\
57355.1 & \textit{XMM} &&&&&&&& *$<44.11$ \\
58977.5 & \textit{SRG}/eROSITA & 1.2 & --- & $52^{+7}_{-6}$ & \textbf{37/40} & --- & $5.2^{+0.7}_{-0.6}$ & 35/40 & *$43.63^{+0.10}_{-0.11}$\\
60249.2 & \textit{XMM} &&&&&&&& *$<44.08$ \\
60410.2 & \textit{Swift}/XRT &&&&&&&& *$<42.93$ \\
\hline
  \end{tabular}
 \end{table*}

\begin{figure*}
\centering
\includegraphics[width=0.33\textwidth]{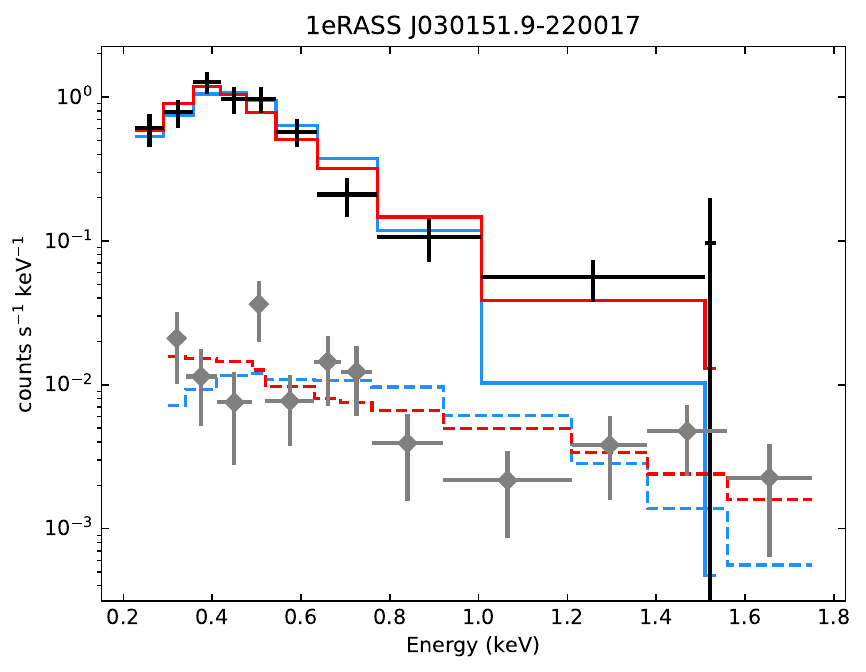}
\includegraphics[width=0.33\textwidth]{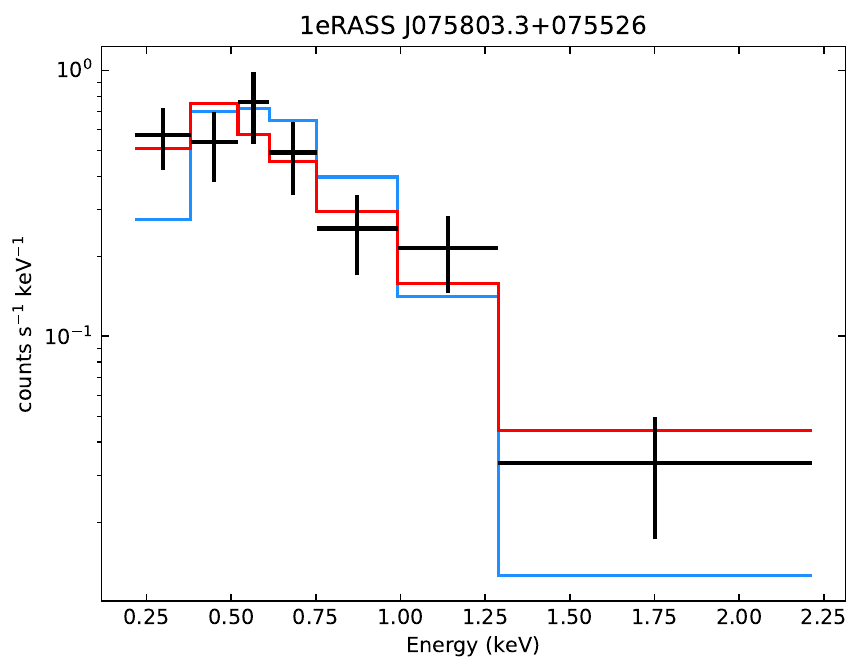}
\includegraphics[width=0.33\textwidth]{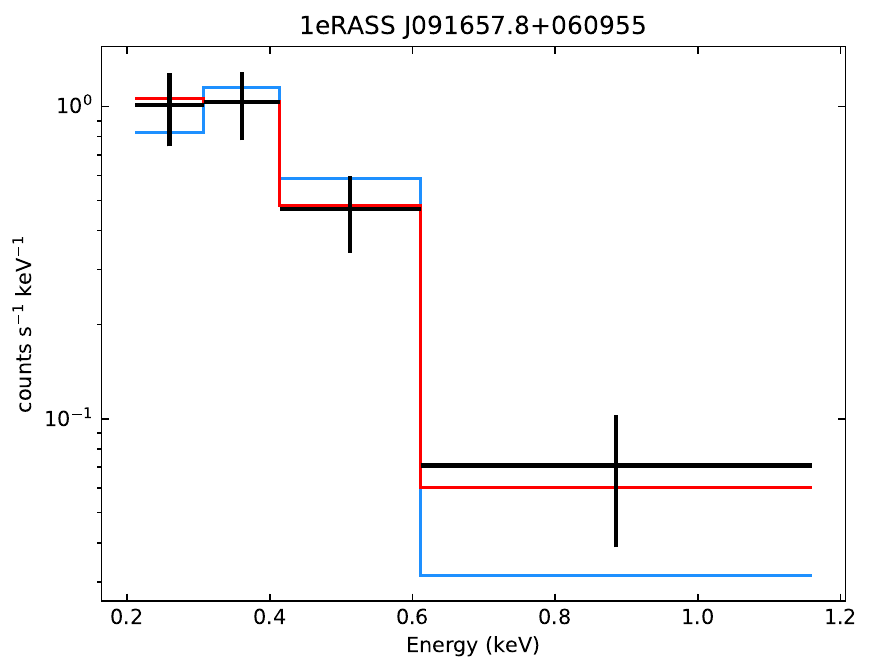}
\includegraphics[width=0.33\textwidth]{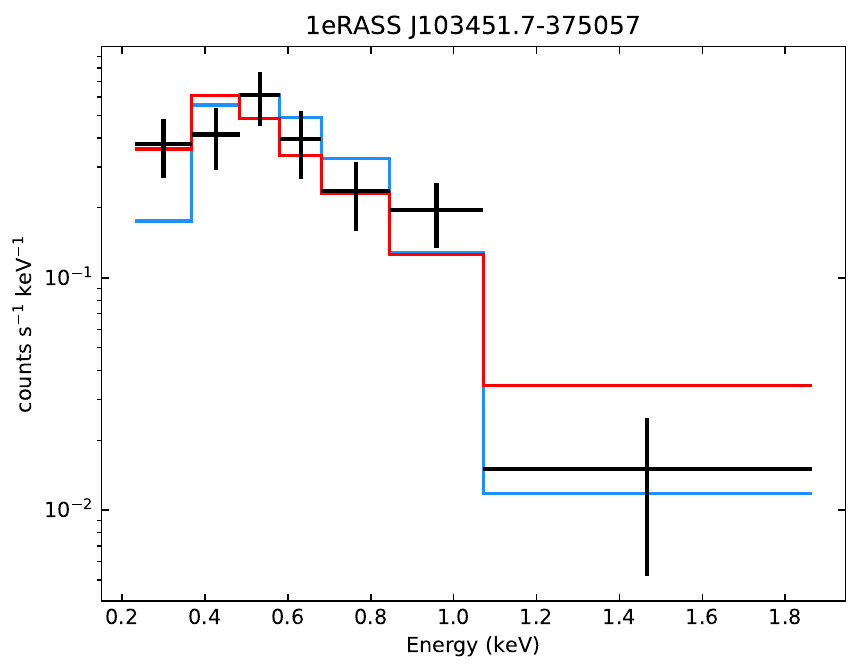}
\includegraphics[width=0.33\textwidth]{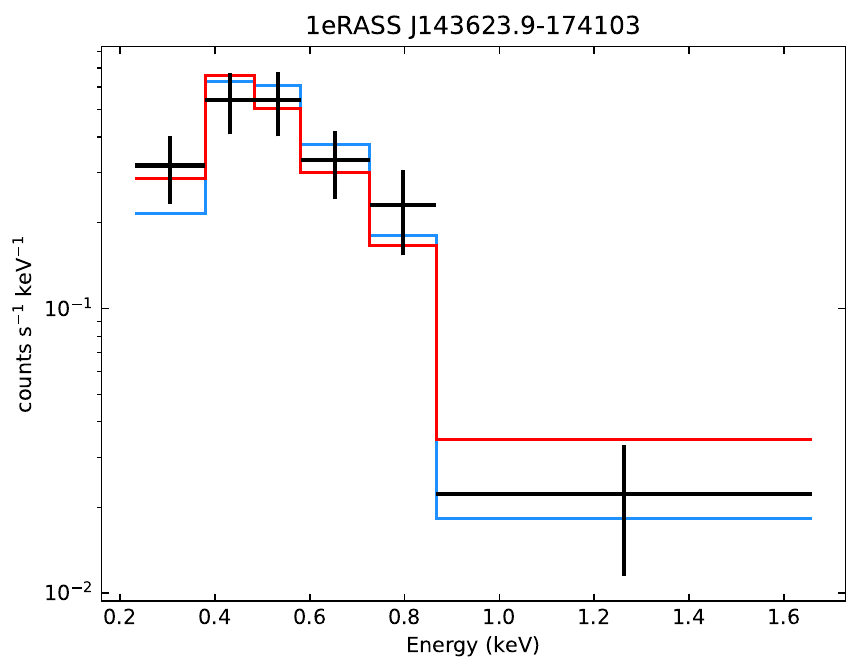}
\includegraphics[width=0.33\textwidth]{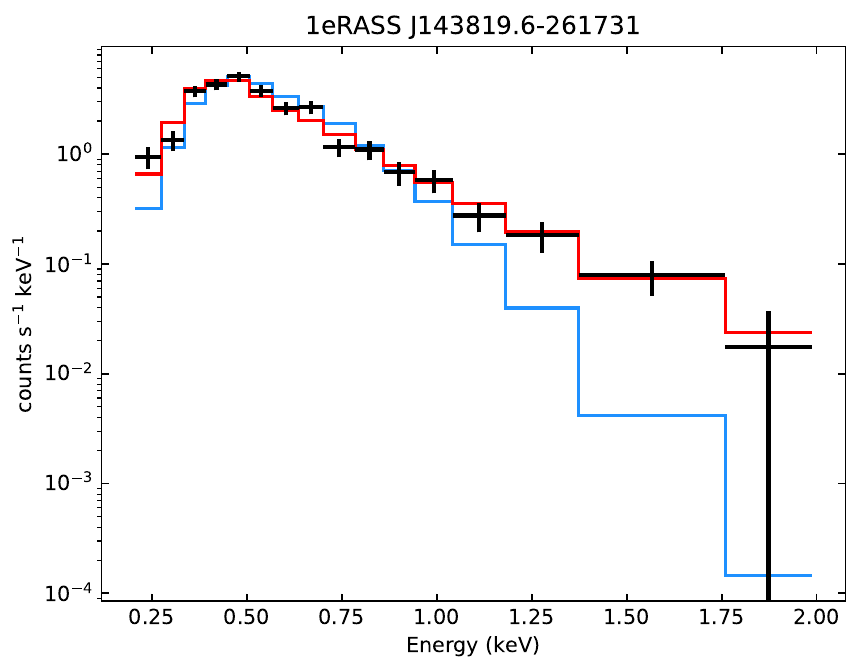}
\includegraphics[width=0.33\textwidth]{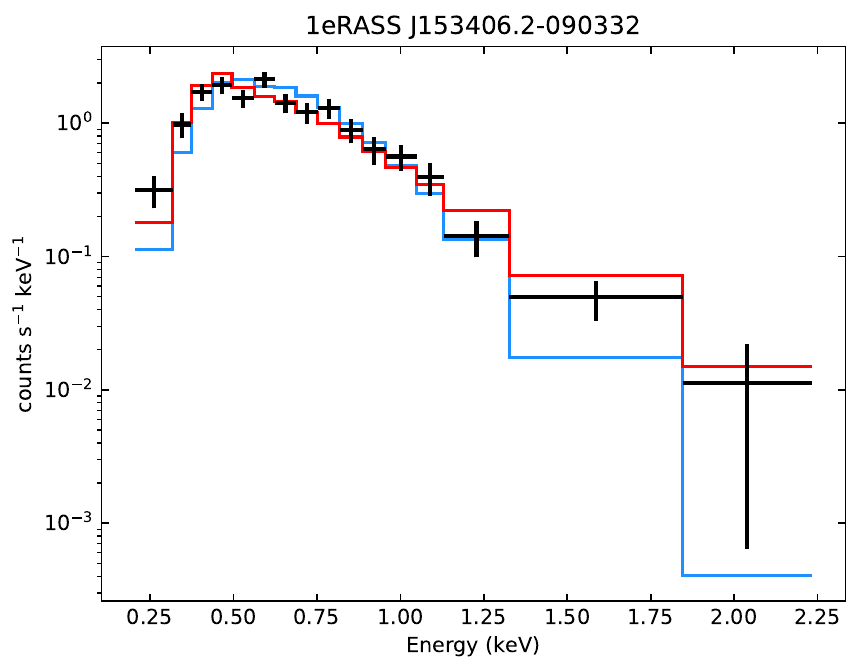}
\includegraphics[width=0.33\textwidth]{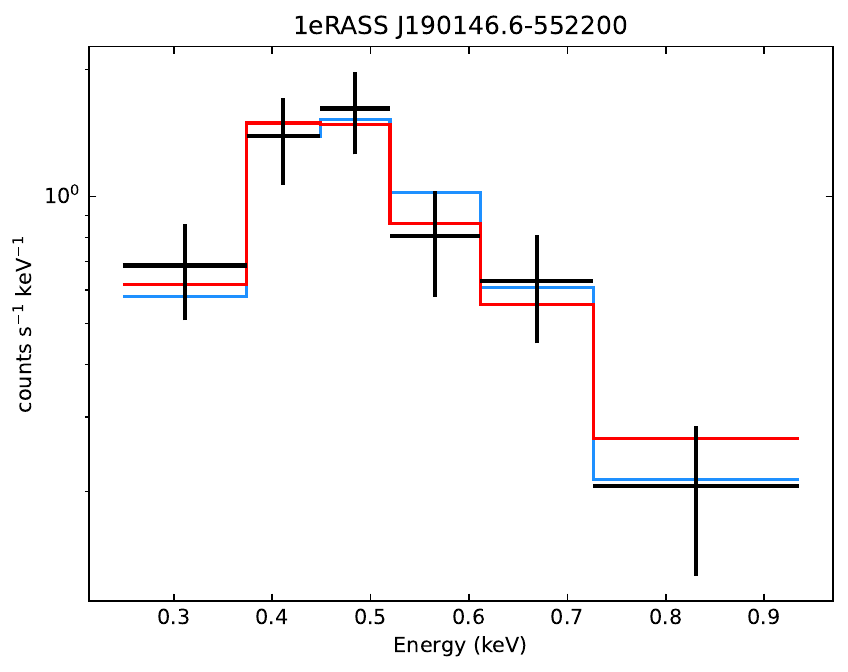}
\includegraphics[width=0.33\textwidth]{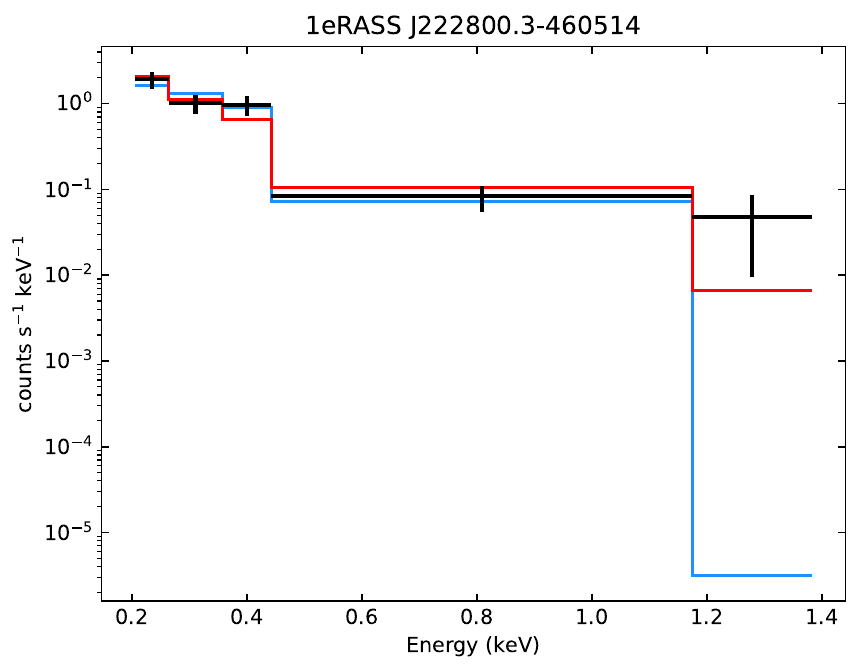}
\caption{The eROSITA spectra of our candidates. In each panel, the spectra are plotted rebinned to 10-$\sigma$ signficance for clarity and the blue and red indicate blackbody and power law fits respectively. In the panel for 1eRASS J0301, we also plot the \textit{Swift}/XRT spectrum observed 1775 days later in grey and the fits to those data with dashed lines. Note that the y-axis gives the instrument specific count rate and therefore the flux of these spectra cannot be directly compared.}
\label{fig:spectra}
\end{figure*}

\section{UVOT photometry}

\begin{table*}
    \caption{The UVOT photometry for our sources. Detections are to 5-$\sigma$ unless noted.}
    \centering
    \begin{tabular}{cccccc}
    \hline
     TDE candidate & MJD & Days since eROSITA-DE detection & Filter & Exposure time (s) & AB magnitude\\
    \hline
1eRASS J0301 & 60364.2 & 1493.9 &\textit{uvm2} & 381.3 & $20.69\pm0.25^a$ \\
& 60645.4 & 1785.1 & \textit{uvw1} & 4505.6 & $19.91\pm0.04$ \\\hline
1eRASS J0758 & 60702.3 & 1740.6 &\textit{uvm2} & 653.7 & $20.16\pm0.12$ \\
& 60705.4 & 1742.7 & \textit{uvw2} & 837.4 & $19.99\pm0.08$ \\\hline
1eRASS J0916 & 60368.8 & 1676.1 & \textit{uvm2} & 812.6 & $>22.09$ \\\hline
1eRASS J1034 & 60702.9 & 1695.1 &\textit{uvm2} & 527.8 & $19.88\pm0.11$ \\
& 60705.5 & 1697.7 & \textit{uvw2} & 531.3 & $19.99\pm0.10$ \\
& 60709.5 & 1701.7 & \textit{uvw2} & 833.2 & $20.28\pm0.10$ \\\hline
1eRASS J1436 & 60368.0 & 1486.9 & \textit{uvm2} & 1098.4 & $>22.24$ \\\hline
1eRASS J1438 & 60362.0 & 1476.2 & \textit{uvm2} & 198.4 & $>20.93$ \\
 & 60367.9 & 1482.1 & \textit{uvm2} & 905.5 & $21.79\pm0.29^b$ \\\hline
 1eRASS J1534 & 60362.8 & 1462.6 &  \textit{uvm2} & 665.2 & $>21.71$ \\
& 60367.1 & 1466.9 & \textit{uvm2} & 750.8 & $>21.77$ \\\hline
1eRASS J1901 & 60364.2 & 1415.6 & \textit{uvm2} & 1035.5 & $21.14\pm0.18$ \\
& 60364.7 & 1416.1 & \textit{uvm2} & 512.6 & $21.31\pm0.31^c$ \\\hline
1eRASS J2228 & 60410.2 & 1432.8 & \textit{uvm2} & 1853.8 & $19.24\pm0.05$ \\
    \hline
    \multicolumn{5}{l}{$^a$Detected to 4.39-$\sigma$; $^b$detected to 3.74-$\sigma$; $^c$detected to 3.55-$\sigma$.}\\
    \end{tabular}
    \label{tab:uvot_photometry}   
\end{table*}

% Don't change these lines
\bsp	% typesetting comment
\label{lastpage}
\end{document}